\documentclass{amsproc}
\usepackage{amsmath, amsthm, amssymb, slashed, framed, amscd}
\usepackage{ifpdf}
\ifpdf
  \usepackage[pdftex]{graphicx}
  \usepackage{epstopdf}
\else
  \usepackage[dvips]{graphicx}
\fi
\usepackage{url}
\newtheorem{theorem}{Theorem}[section]

\theoremstyle{definition}

\newtheorem{example}[theorem]{Example}

\theoremstyle{remark}

\newdimen\tableauside\tableauside=1.0ex
\newdimen\tableaurule\tableaurule=0.4pt
\newdimen\tableaustep
\def\phantomhrule#1{\hbox{\vbox to0pt{\hrule height\tableaurule width#1\vss}}}
\def\phantomvrule#1{\vbox{\hbox to0pt{\vrule width\tableaurule height#1\hss}}}
\def\sqr{\vbox{%
\phantomhrule\tableaustep
\hbox{\phantomvrule\tableaustep\kern\tableaustep\phantomvrule\tableaustep}%
\hbox{\vbox{\phantomhrule\tableauside}\kern-\tableaurule}}}
\def\squares#1{\hbox{\count0=#1\noindent\loop\sqr
\advance\count0 by-1 \ifnum\count0>0\repeat}}
\def\tableau#1{\vcenter{\offinterlineskip
\tableaustep=\tableauside\advance\tableaustep by-\tableaurule
\kern\normallineskip\hbox
    {\kern\normallineskip\vbox
      {\gettableau#1 0 }%
     \kern\normallineskip\kern\tableaurule}%
  \kern\normallineskip\kern\tableaurule}}
\def\gettableau#1 {\ifnum#1=0\let\next=\null\else
  \squares{#1}\let\next=\gettableau\fi\next}

\numberwithin{equation}{section}

\title{VOA$[M_4]$}

\author{Boris Feigin}
\address{National Research University Higher School of Economics,
Russian Federation, International Laboratory of Representation Theory\\
\newline
Mathematical Physics, Russia, Moscow, 101000, Myasnitskaya ul., 20\\
\newline
Landau Institute for Theoretical Physics, Russia, Chernogolovka, 142432, pr.Akademika Semenova, 1a.}
\email{borfeigin@gmail.com}
\author{Sergei Gukov}
\address{California Institute of Technology, Pasadena, CA 91125, USA\\
\newline
Max-Planck-Institut f\"ur Mathematik, Vivatsgasse 7, D-53111 Bonn, Germany.}
\email{gukov@theory.caltech.edu}

\font\teneurm=eurm10 \font\seveneurm=eurm7 \font\fiveeurm=eurm5
\newfam\eurmfam
\textfont\eurmfam=\teneurm \scriptfont\eurmfam=\seveneurm
\scriptscriptfont\eurmfam=\fiveeurm

 \font\teneusm=eusm10 \font\seveneusm=eusm7 \font\fiveeusm=eusm5
\newfam\eusmfam
\textfont\eusmfam=\teneusm \scriptfont\eusmfam=\seveneusm
\scriptscriptfont\eusmfam=\fiveeusm

\font\tencmmib=cmmib10 \skewchar\tencmmib='177
\font\sevencmmib=cmmib7 \skewchar\sevencmmib='177
\font\fivecmmib=cmmib5 \skewchar\fivecmmib='177
\newfam\cmmibfam
\textfont\cmmibfam=\tencmmib \scriptfont\cmmibfam=\sevencmmib
\scriptscriptfont\cmmibfam=\fivecmmib

\def\example#1{\bgroup\narrower%\footnotefont
\baselineskip\footskip\bigbreak
\hrule\medskip\nobreak\noindent {\bf Example}. {\it #1\/}\par\nobreak}
\def\endexample{\medskip\nobreak\hrule\bigbreak\egroup}

\newcommand{\be}{\begin{equation}}
\newcommand{\ee}{\end{equation}}
\newcommand{\bea}{\begin{eqnarray}}
\newcommand{\eea}{\end{eqnarray}}

\newcommand{\unknot}{{\,\raisebox{+.08cm}{\includegraphics[width=.4cm]{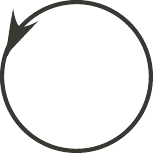}}\,}}
\newcommand{\C}{\mathbb{C}}
\newcommand{\Z}{\mathbb{Z}}
\newcommand{\R}{\mathbb{R}}

\def\frak{\mathfrak}

\def\hat{\widehat}
\def\bar{\overline}

\def\CA{{\mathcal A}}
\def\CB{{\mathcal B}}
\def\CC{{\mathcal C}}
\def\CE{{\mathcal E}}

\def\CL{{\mathcal L}}
\def\CM{{\mathcal M}}
\def\CN{{\mathcal N}}
\def\CO{{\mathcal O}}

\def\CR{{\mathcal R}}

\def\CT{{\mathcal T}}
\def\CU{{\mathcal U}}
\def\CV{{\mathcal V}}
\def\CW{{\mathcal W}}

\def\unknot{{\,\raisebox{-.08cm}{\includegraphics[width=.4cm]{unknot}}\,}}

\def\cp{{\mathbb{C}}{\mathbf{P}}}

\DeclareMathOperator{\Tr}{Tr}
\DeclareMathOperator{\tr}{tr}

\begin{document}

%    General info

\begin{abstract}
We take a peek at a general program that associates vertex (or, chiral) algebras to smooth 4-manifolds in such a way
that operations on algebras mirror gluing operations on 4-manifolds and, furthermore,
equivalent constructions of 4-manifolds give rise to equivalences (dualities) of the corresponding algebras.
\end{abstract}

\maketitle

\tableofcontents
%\newpage
%\vspace*{20mm}

%%%%%%%%%%%%%%%%%%%%%%%%%%%%%%%%%%%%%%%%%%%%%%%%%%%%%%%%%%%%%%%%%%%%%%%%%%%%%%%%%%%%

\section{Introduction}
\label{sec:intro}

The main goal of this paper is to blaze new trails between topology and algebra, in particular, topology of 4-manifolds and vertex operator algebra. While various connections between these two subjects emerged during the past 2-3 decades, the questions of prime interest in topology --- that involve (cutting and gluing) surgery operations --- remain, surprisingly, untouched.
Similarly, from the algebra perspective, many traditional connections to topology rely on algebraic structures used as an input, {\it i.g.} Frobenius algebras in the construction of 2d TQFTs or modular tensor categories in the construction of 3d TQFTs. More recent developments, however, suggest that such algebraic structures can themselves be topological invariants of manifolds, so that one can speak of $\text{VOA} [M_4]$ or $\text{MTC} [M_3]$.

Pointing a flashlight directly at these underappreciated aspects can potentially be very beneficial for the development of both subjects.
Specifically, we describe a large class of vertex operator algebras (VOAs),
each labeled by a choice of a smooth 4-manifold $M_4$ and a root system $\frak g$
of ADE Cartan type.\footnote{More often than not, our VOAs will, in fact, be {\it conformal} vertex
algebras and/or vertex {\it superalgebras}. However, to avoid using too verbose language,
we will not explicitly emphasize such additional structures, unless they become center of our attention.}
We focus on algebraic operations that correspond to cutting and gluing operations in topology and use equivalences of VOAs that correspond to different ways of building the same 4-manifold as tests of the proposed dictionary.

Mathematically, the problem we wish to tackle is motivated by studying the algebraic structure that one may find
on cohomology (or K-theory) of the moduli spaces of $G$-instantons on $M_4$, where $G$ is a compact Lie group with ${\frak g} = \text{Lie} (G)$:
\be
\text{VOA} [M_4] ~~~\rotatebox[origin=c]{-90}{$\circlearrowright$}~~~ V_{\rho} := \oplus_n H^* \left( \CM_n (M_4, G, \rho) \right)
\label{VOAmoduli}
\ee
It aims to generalize Nakajima's construction for ALE spaces \cite{Nakajima} to arbitrary 4-manifolds. In particular, when $M_4$ is a complex surface, the algebra $\text{VOA} [M_4]$ can be thought of as generated by Hecke modifications of a gauge bundle (sheaf) $\CE$ along curves $C_i \subset M_4$:
\be
0 \; \longrightarrow \; \CE' \; \longrightarrow \; \CE  \; \longrightarrow \; O(-1) \vert_{C_i} \; \longrightarrow \; 0
\ee
In order to emphasize the dependence of $\text{VOA} [M_4]$ on the choice of the root system,
sometimes we write it either as $\text{VOA} [M_4,G]$ or $\text{VOA} [M_4,{\frak g}]$,
and use a shorter notation when this additional data is fixed and clear from the context.
In this paper we mostly focus on rank-1 case, relegating the detailed discussion
of higher-rank $G$ to our next paper.\footnote{For $G=U(1)$ and arbitrary $M_4$,
the complete description of $\text{VOA} [M_4,G]$ can be easily extracted from \cite{Gadde:2013sca,Dedushenko:2017tdw}.}

There are at least three ingredients in \eqref{VOAmoduli} that call for immediate attention and will be progressively
improved throughout the text, to some extent, even below in this section: the moduli space $\CM_n$, its cohomology, and the extra data $\rho$.
As for the first ingredient, the most natural candidate for $\CM_n$, namely the moduli space of instantons on $M_4$ with $c_2 (\CE) = n$,
is a non-compact singular space which is rather sensitive to metric $g(M_4)$ that makes it hard to find a suitable
cohomology theory that would actually be invariant.
A slightly better candidate, motivated by physics, is the moduli space of solutions to Vafa-Witten equations \cite{Vafa:1994tf} on $M_4$.
(Part of the reason it is a better-behaved moduli space has to do with
the fact that Vafa-Witten theory is an example of {\it balanced} topological theory \cite{Dijkgraaf:1996tz}.)

Similarly, one needs a suitable cohomology theory, to which K-theory will often provide a first approximation.
A better candidate, though, is the Floer homology based on
topological twist of 5d super-Yang-Mills theory \cite{Haydys:2010dv,Witten:2011zz}
that categorifies (numerical) Vafa-Witten invariants of $M_4$. The Floer complex in this five-dimensional theory is constructed by studying solutions to the following partial differential equations (PDEs) on $\R \times M_4$:
\be
\begin{aligned}
F^+ - \frac{1}{4} B \times B - \frac{1}{2} D_t B & = 0 \\
F_{t \mu} + D^{\nu} B_{\nu \mu} & = 0
\end{aligned}
\qquad \text{where} \qquad
\begin{aligned}
A & \in \CA (\CE) \\
B & \in \Omega^2_+ (\text{ad} \CE)
\end{aligned}
\label{HWpdes}
\ee
where $\R$ is parametrized by the ``time'' coordinate $t$,
$F$ is the curvature of the gauge connection $A$, and $B$ is a self-dual 2-form on $M_4$. When $A$ and $B$ are independent of $t$, these equations reduce to the Vafa-Witten equations on $M_4$; the $t$-component of $A$ then becomes an $\text{ad} \CE$-valued scalar on $M_4$ denoted by $C$ in \cite{Vafa:1994tf}.
Following Floer's construction \cite{Floer1988}, the goal is to use such ``stationary'' solutions as generators of the complex and to define a homology with respect to the differential that ``counts'' $t$-dependent solutions, the ``instantons.'' Since the moduli spaces involved in this problem are non-compact and always singular, carrying this out in practice requires substantial work which is currently underway, see {\it e.g.} \cite{Taubes:2017pzr,He:2017iyg,Leung:2018ofs,Tanaka}.

Finally, $\rho: \pi_1 (\partial M_4) \to G_{\C}$ in \eqref{VOAmoduli}
labels the choice of a boundary condition when $M_4$ is non-compact.
Indeed, boundary conditions in Vafa-Witten theory are naturally labeled by complex flat connections on $M_3 = \partial M_4$.
However, as we shall see below, it is better to think of $\rho$ as an element in $K^0 \big( \text{MTC} [M_3] \big)$,
the Grothendieck ring of the (modular) tensor category $\text{MTC} [M_3]$ introduced in \cite{Gukov:2016gkn}.\\

\subsection{Transgression and QFT-valued topological invariants}

Physically, $\text{VOA} [M_4]$ is the (left-moving) chiral algebra of
a 2d $\CN=(0,2)$ superconformal theory $T[M_4]$ introduced in \cite{Gadde:2013sca}
and obtained via compactification of the six-dimensional $(0,2)$ fivebrane theory on a 4-manifold $M_4$.

The basic idea of compactification and assigning field theories to manifolds is analogous to
the operations of fiber integration or push-forward in mathematics.
In order to illustrate this parallel, consider a fully extended $d$-dimensional TQFT which,
according to the (extension of) standard Atiyah-Segal axioms, assigns a number to a closed $d$-manifold,
a vector space to a closed $(d-1)$-manifold, and so on, all the way up to a $(d-1)$-category assigned to a point.
In this framework, it is rather clear that any closed $n$-dimensional manifold $M_n$, with $n \le d$,
defines a TQFT in dimension $d-n$ ({\it i.e.} a functor from Bord$_{d-n}$) by a ``push-forward''
\be
\text{TQFT}_{d-n} \big( \ldots \big) \; := \; \text{TQFT}_{d} \big( \, \ldots \times M_n \big)
\label{TQFTreduction}
\ee
which is analogous to evaluating cohomology classes of a (trivial) bundle $M_n \times (\ldots)$ along the fiber $M_n$.
In other words, if we wish to know what the functor $\text{TQFT}_{d-n}$ assigns to a space from Bord$_{d-n}$
represented by ellipsis in \eqref{TQFTreduction}, we can simply take a product of that space with $M_n$ and apply $\text{TQFT}_{d}$.

Non-trivial topological theories in high dimensions are quite rare.
However, many higher-dimensional supersymmetric theories can be made \cite{Witten:1988ze,Bershadsky:1995qy}
at least ``partly topological'' on spaces up to a certain dimension $n$.
In such cases, a version of \eqref{TQFTreduction} still applies and defines a topological invariant of $M_n$,
albeit valued in QFTs rather than in TQFTs. The QFTs relevant to us here are all conformal,
both in six dimensions and also in two dimensions\footnote{Conformal field theories (CFTs)
are scale-invariant end-points of the RG flow.
If theories $T[M_4]$ were not defined as conformal fixed points of the RG flow,
they would depend on scale and possibly other details of the compactification on $M_4$.}
after (path) integrating along $M_4$ fibers:
\be
2|2\text{-CFT} \big( \ldots \big) \; := \; 6|16\text{-CFT} \big( \, \ldots \times M_4 \big)
\label{CFTreduction}
\ee
Although the six-dimensional theory here can not be made fully topological, it admits a holomorphic
twist along 2-dimensional surface $\Sigma$ (in addition to the topological twist along the 4-manifold $M_4$)
which leads to chiral algebra $\text{VOA} [M_4]$.
In M-theory, the corresponding geometry looks like
\begin{equation}
 \begin{array}{cccc}
11\text{d~space-time:} & \quad T^* \Sigma & ~\times & \Lambda^{2,+} (M_4) \\
   & \quad \cup &  & \cup \\
N\,\text{fivebranes:} & \quad \Sigma & ~\times & M_4 \\
 & \quad \overbrace{\footnotesize{\text{holomorphic twist}}} & ~~ & \overbrace{\footnotesize{\text{topological twist}}} \\
 \end{array}
\label{TM4def}
\end{equation}
Holomorphic twists of two-dimensional $(0,2)$ theories, of which $T[M_4]$ are examples, have a rather long history \cite{Silverstein:1995re,Katz:2004nn,Dedushenko:2015opz}, and generalizations to holomorphic twists of higher-dimensional QFTs go back to the work of Johansen \cite{Johansen:1994aw,Johansen:1994ij,Johansen:1994ud} in mid-90s.

This way of associating a 2d superconformal theory \eqref{CFTreduction}
to a 4-manifold is similar to the so-called 3d-3d correspondence
that associates a $3|4$-CFT called $T[M_3]$ to a 3-manifold $M_3$ (and a choice of root system).
In fact, 3d-3d correspondence and the theory $T[M_3]$ will play a role in our story as well when we consider gluing 4-manifolds along $M_3$.
Both $T[M_4]$ and $T[M_3]$ can (and should!) be viewed as topological invariants valued in quantum field theories, from
which more traditional topological invariants can be extracted by evaluating various partition functions of these theories,
studying their Hilbert spaces, {\it etc.}

Much like $\text{VOA} [M_4]$ captures the ``mathematical content'' of the 2d $\CN=(0,2)$ superconformal theory $T[M_4]$,
the modular tensor category $\text{MTC} [M_3]$ mentioned earlier captures the mathematical content of $T[M_3]$ relevant to gluing operations.
Various constructions of $T[M_4]$ and $T[M_3]$ that will be our main focus here are very much in line with
the general ideas of BPS/CFT correspondence formulated circa 2002-2004 \cite{Nikitatalk} (see also introduction to \cite{Nekrasov:2015wsu}),
which we hope to enrich by connecting the latter to dualities of 2d $\CN=(0,2)$ and 3d $\CN=2$ theories.
In particular, one of our main goals is to describe the behavior of $\text{VOA}[M_4]$ under
cutting and gluing operations on $M_4$ and to study equivalences
of VOA's associated to different constructions of the same 4-manifold.

Another goal of the present paper is to ``connect the dots,'' {\it i.e.} to bring together various developments where
similar vertex algebras appear in closely related physical systems, but have not yet been explicitly connected.
Thus, some of the vertex algebras we are going to mention are familiar in the context of
the BPS/CFT and AGT correspondence \cite{Belavin:2011sw,Bonelli:2012ny,Bershtein:2013oka,Bruzzo:2013daa,Bershtein:2016mxz}
and one novelty of the present approach
is to interpret even these familiar examples as chiral (left-moving) algebras of 2d superconformal theories $T[M_4]$.
Needless to say, it is important to test this proposed interpretation by studying theories $T[M_4]$ and verifying that their chiral
algebras indeed agree with VOAs discussed here.\\

The way we presented mathematical and physical motivations here, makes it non-obvious that they actually describe the same problem.
The relation becomes more apparent if, in the physical setup \eqref{TM4def}, we take $\Sigma = T^2$ and consider two
different orders of compactification:
$$
\begin{array}{ccccc}
\; & \; & \text{6d $(2,0)$ theory} & \; & \; \\
\; & \; & \text{on $T^2 \times M_4$} & \; & \; \\ \; & \swarrow & \; & \searrow & \; \\
\text{$4d$ Vafa-Witten theory} & \; & \; & \; & \text{2d $(0,2)$ theory $T[M_4]$} \\
\text{on $M_4$} & \; & \; & \; & \text{on $T^2$}
\end{array}
$$
both of which describe the same partition function of the 6d superconformal theory on $T^2 \times M_4$ and, therefore,
must yield the same result.
Indeed, on the one hand, the partition function of Vafa-Witten theory on $M_4$ is a ``decategorification'' of \eqref{VOAmoduli},
{\it i.e.} the graded Euler characteristic of $\cup_n \CM_n$.
And, on the other hand, the elliptic genus of a 2d $\CN = (0,2)$ superconformal theory $T[M_4]$
is a character of the left-moving chiral algebra,
which from now on will be our main working definition of $\text{VOA} [M_4]$:
\be
\chi_{\rho} (q) \; := \; \Tr_{V_{\rho}} q^{L_0 - \frac{c}{24}} \; = \; Z_{VW} (M_4,\rho;q)
\label{ZVWchar}
\ee
Recall, that a general 2d $(0,2)$ superconformal theory comes equipped with its left and right chiral algebras, which include $\CN=0$ Virasoro and $\CN=2$ super-Virasoro subalgebras, respectively, with central charges $c_L$ and $c_R$. The $\CN=2$ algebra in the right-moving sector also includes a supercharge $\bar Q_+$, such that only states in $\bar Q_+$ cohomology contribute to the elliptic genus~\cite{Witten:1986bf}. As a result, one can define the elliptic genus of a 2d $(0,2)$ superconformal theory either by taking the trace over all states or, equivalently, over the states in $\bar Q_+$ cohomology. Denoting the latter by $V$, or $V_{\rho}$ in the context of 2d $\CN = (0,2)$ theory $T[M_4]$, we arrive at the identification \eqref{ZVWchar}.
Note, since left and right chiral algebras commute, $V_{\rho}$ carries the action of the left chiral algebra that we call $\text{VOA} [M_4]$.
\\

\subsection{Useful tools}

The physical formulation \eqref{TM4def} provides at least three pieces of
valuable information about our vertex operator algebra $\text{VOA} [M_4]$,
which are relatively accessible for general 4-manifolds (ordered here from the simpler to more sophisticated):

\begin{itemize}

\item
the central charge $c_L$ of $\text{VOA} [M_4]$ or, equivalently, the central charge
of the left-moving sector of the 2d $\CN=(0,2)$ superconformal theory $T[M_4]$;

\item
the information about the fusion product and the modular $SL(2,\Z)$ action
on the representation (sub)ring of $\text{VOA} [M_4]$ when $M_4$ has boundary;

\item
the characters of its modules \eqref{ZVWchar}.

\end{itemize}

\noindent
Let us briefly comment on each of the three items.

The central charge of $\text{VOA} [M_4]$ is inherited from the full superconformal theory $T[M_4]$
that contains both left- and right-moving sectors.
For example, for $G=SU(2)$ and general $M_4$ (of irreducible holonomy), we have\footnote{\label{foot:holonomy}When
the holonomy of $M_4$ is reduced, this formula may no longer apply. For example, the values of $c_L$
listed in Table~\ref{table:cLcR} for $M_4 = \cp^1 \times \Sigma_{g,n}$ are taken from \cite{Putrov:2015jpa}
and differ from \eqref{cLgeneric}. For further discussion and references see {\it e.g.} \cite{Dedushenko:2017tdw}.}
\be
c_L \; = \; 13 \chi + 18 \sigma
\label{cLgeneric}
\ee
where $\chi$ and $\sigma$ are, respectively, the Euler characteristic and the signature of $M_4$.
Simple examples of 4-manifolds and the corresponding values of $c_L$ are listed in Table~\ref{table:cLcR}.

\begin{table}\begin{center}
\begin{tabular}{|c|c|c|}
\hline
\rule{0pt}{5mm}
$M_4$ & $c_L$ & $c_R$ \\[3pt]
\hline
\hline
\rule{0pt}{5mm}
$S^4$ & $26 = 2 + 24$ & $27 = 3 + 24$ \\[3pt]
\hline
\rule{0pt}{5mm}
$\cp^2$ & $57$ & $60$ \\[3pt]
\hline
\rule{0pt}{5mm}
$\bar \cp^2$ & $21$ & $21$ \\[3pt]
\hline
\rule{0pt}{5mm}
$S^2 \times S^2$ & $52$ & $54$ \\[3pt]
\hline
\rule{0pt}{5mm}
$m \cp^2 \# n \bar \cp^2$ & $26 + 31m - 5n$ & $27 + 33 m - 6n$ \\[3pt]
\hline
\rule{0pt}{5mm}
$\cp^1 \times \Sigma_{g,n}$ & $2g + 4n + 4$ & $6n+6$ \\[3pt]
\hline
\rule{0pt}{5mm}
K3 & $24$ & $12$ \\[3pt]
\hline
\end{tabular}\end{center}
\caption{Central charges of the 2d $\CN=(0,2)$ superconformal theory $T[M_4,G]$ with $G=SU(2)$ and various $M_4$.
Of particular interest to us is the value $c_L$, which is the central charge of the vertex operator algebra $\text{VOA} [M_4]$.}
\label{table:cLcR}
\end{table}

The second useful tool, $\text{MTC} [M_3]$, plays a role only in study of 4-manifolds with boundary and in questions related to
gluing along those boundaries, as was already briefly mentioned earlier.
Specifically, there is a map \cite[eq.(2.49)]{Gukov:2016gkn} from objects of $\text{MTC} [M_3]$ to modules of $\text{VOA} [M_4]$ that, from the viewpoint of 5d gauge theory on $\R \times M_4$, assigns boundary conditions to objects of $\text{MTC} [M_3]$. And, $\text{MTC} [M_3]$ itself is the (sub)category of representations of our chiral algebra $\text{VOA} [M_4]$.
In particular, its Grothendieck ring is (a subring of) the representation ring of $\text{VOA} [M_4]$:
\be
K^0 \big( \text{MTC} [M_3] \big) \quad \subseteq \quad \text{Representation ring of VOA$[M_4]$}
\label{Krepring}
\ee
Following \cite{Gukov:2016gkn}, it can be computed for rather general 3-manifolds.
%(Explicit examples of such computations will be presented below.)
Concretely, when all $G_{\C}$ flat connections on $M_3$ are isolated,
as {\it e.g.} for the Poincar\'e sphere $P = -\Sigma (2,3,5)$ or more general plumbings discussed below,
we expect simple objects of $\text{MTC} [M_3]$ to be in one-to-one correspondence with representations
\be
\rho: \quad \pi_1 (M_3) \; \to \; G_{\C}
\label{rhorep}
\ee
modulo conjugation.
For example, when $G=U(N)$ and $M_3 = L(p,1)$ is the Lens space, we have
\be
K^0 \big( \text{MTC} [L(p,1)] \big) \; = \; \text{Verlinde algebra of $\hat{\frak{su}(p)}_N$}
\ee
To give another example, from a different family, consider a 3-manifold defined
by the following plumbing graph\footnote{What a plumbed 3-manifold is and how to associate to it
a modular tensor category will be explained further in the main text.}
\be
\begin{array}{ccc}
& \overset{\displaystyle{-2}}{\bullet} & \\
& \vline & \\
\overset{\displaystyle{-2}}{\bullet}
\frac{\phantom{xxxx}}{\phantom{xxxx}}
& \underset{\displaystyle{-2}}{\bullet} &
\frac{\phantom{xxxx}}{\phantom{xxxx}}
\overset{\displaystyle{-2}}{\bullet}
\end{array}
\qquad : \qquad
\text{MTC} [M_3] \; = \; \text{3-fermion model}
\label{MTCD4}
\ee
for which $\text{MTC} [M_3,U(1)]$ is the modular tensor category that governs fusion and braiding in the so-called
``3-fermion model,'' a close cousin of Kitaev's toric code model,
also widely used in the literature on symmetry protected topological (SPT) phases of matter.

Physically, $\text{MTC} [M_3,G]$ can be thought of in several rather different but equivalent ways. On the one hand, it can be defined as a category of line operators in 3d $\CN=2$ theory $T[M_3,G]$. On the other hand, it encodes in a convenient way the information about all topologically twisted partition functions (``indices'') of 3d $\CN=2$ theory $T[M_3,G]$ on $S^1 \times \Sigma$.
In particular, the former description of $\text{MTC} [M_3]$ is especially useful for understanding its braiding and fusion properties~\cite{in progress}, while the latter involves $S$ and $T$ matrices of $\text{MTC} [M_3,G]$ in a crucial way that are related to the modular transformations of the VOA characters \eqref{ZVWchar} for $M_4$ bounded by $M_3$, {\it cf.} \eqref{Krepring}.

%Thus, for the Poincar\'e sphere $P = -\Sigma (2,3,5)$ and $G = SU(2)$,
%this gives three simple objects with the fusion algebra of the Ising model.
%For example, if $G = SU(2)$ and $M_3$ is the Poincar\'e sphere $P = -\Sigma (2,3,5)$, we find
%\be
%K^0 \big( \text{MTC} [- \Sigma (2,3,5)] \big) \; = \; \text{Verlinde algebra of $( \hat{E_8} )_2$}
%\ee
%It gives the same fusion algebra as in the most familiar 2d conformal field theory, namely the Ising model,
%generated by three operators $1$, $\sigma$, and $\epsilon$ that correspond to three $SU(2)$ flat connections on $M_3 = - \Sigma (2,3,5)$.

Finally, the character \eqref{ZVWchar} is equal to the Vafa-Witten partition function on $M_4$ \cite{Vafa:1994tf}.
While this ingredient provides most of the information about the $\text{VOA}[M_4]$, it is also the one most difficult to compute.
(See, however, \cite{Tanaka:2017jom,Gottsche:2017vxs} for recent progress in this area.)
When $M_4$ has boundary, \eqref{ZVWchar} is labeled by a choice of complex flat connection $\rho$ or, when flat connections are isolated, by a generator of the ring $K^0 \big( \text{MTC} [M_3] \big)$.
In such cases, the modular properties of \eqref{ZVWchar} may deviate from the traditional modularity along the lines of \cite{Vafa:1994tf,Alim:2010cf,Cheng:2018vpl}, {\it e.g.} it can be mock modular, mock of higher depth, or possibly even some new and unexplored form of modularity.
The modular properties of \eqref{ZVWchar} for general 4-manifolds bounded by $M_3$ certainly deserve further study.
Even when $M_4$ does not have a boundary, note that $V_{\rho}$ in \eqref{VOAmoduli} does {\it not} need to be a vacuum module.

\subsection{Chiral correlators and 4-manifold invariants}

Yet another motivation for the present work is a new perspective on 4-manifold invariants.
Hopefully, it will stimulate the development of new 4-manifold invariants in the future.

A useful example to keep in mind, especially when things get too technical, is the case of $G=U(1)$.
Not only can it provide intuition about $\text{VOA} [M_4,G]$ with non-abelian $G$,
but it can also help appreciate why vertex algebras are relevant in a first place.
Suppose, for example, one is interested in studying traditional gauge theoretic invariants of $M_4$,
such as Seiberg-Witten invariants. At a first glance, this problem has little to do with Vafa-Witten invariants,
let alone vertex algebras.
A slight variation of the Seiberg-Witten theory, where one has several spinor fields $\Psi_i$, $i=1,\ldots, N_f$:
\begin{eqnarray}
F_A^+ & = & \sum_{i=1}^{N_f} (\Psi_i \bar \Psi_i)^+, \label{NfSWeq} \\
D \!\!\!\! \slash \, \Psi_i & = & 0, \qquad i = 1, \ldots, N_f, \nonumber
\end{eqnarray}
turns out to be a rather challenging problem, even for $G=U(1)$, because the moduli spaces of solutions to
these PDEs are non-compact
(and so the integrals over these moduli spaces are not defined) \cite{MR1392667,MR3432158,2016arXiv160701763H}.
Luckily, in this case, non-compactness can be easily ``cured'' by working equivariantly
with respect to $SU(N_f)$ symmetry that acts on $\Psi_i$ in an obvious way.
It is easy to see that the fixed point sets of this action are compact \cite{Dedushenko:2017tdw}.
Hence, the corresponding equivariant integrals can be defined and the final result is a function of
the equivariant parameters $z_i$, $i=1,\ldots, N_f$.

The way we formulated it, this problem looks like a typical problem in gauge theory and has no obvious connection
to vertex algebras ... until we evaluate the integrals and realize that, as functions of $z_i$, they are equal to
chiral correlators in $\text{VOA} [M_4,G]$ with $G=U(1)$.
Schematically (see \cite{Dedushenko:2017tdw} for details),
\be
SU(N_f)\text{-equivariant} \int_{\CM_{N_f}} (\ldots)
\quad = \quad
\langle \CO (z_1) \ldots \CO (z_{N_f}) \rangle_{\text{VOA} [M_4]}
\label{chiralcorr}
\ee

The surprising appearance of VOA does not stop here.
For $G=U(1)$, the relevant vertex algebra as well as the full physical 2d $\CN=(0,2)$ theory $T[M_4]$
can be easily determined for any 4-manifold $M_4$ using the standard rules of Kaluza-Klein reduction.
(The resulting VOA is essentially a lattice algebra for $H_2 (M_4,\Z)$, modulo an important detail that
will be explained in section~\ref{sec:3manifolds}.)
If we now compute its elliptic genus, we obtain a completely different gauge theoretic invariant,
namely the Vafa-Witten partition function of $M_4$ with $G=U(1)$, {\it cf.} \eqref{ZVWchar}.
This way, $\text{VOA} [M_4]$ serves as a natural home to seemingly different and unrelated problems in gauge theory.

Conversely, %when $G$ is non-abelian,
a computation of Donaldson-Witten type invariants based on \eqref{NfSWeq}
or Vafa-Witten partition functions for a given 4-manifold $M_4$
can help identifying the corresponding algebra $\text{VOA} [M_4]$
by interpreting these invariants as chiral correlators or characters, respectively.

%%%%%%%%%%%%%%%%%%%%%%%%%%%%%%%%%%%%%%%%%%%%%%%%%%%%%%%%%%%%%%%%%%%%%%%%%%%%%%%%%%%%

\section{Gluing via extensions: Toric $M_4$}
\label{sec:extensions}

There are two basic techniques to construct new VOAs:
via extensions or via restrictions to subalgebras by means of screening operators, BRST reduction, and their variants
(see \cite{Feigin:2017edu} for a review and introduction).
Both have applications to vertex algebras associated to 4-manifolds
and describe gluing operations in different setting.
Here we start with basic gluing operations that are described by extensions of the product of two VOA's by their bimodules.
In such constructions, the modules of the resulting algebra are usually obtained via induction functors.

Let $\text{Vir}_b$ denote the Virasoro VOA with central charge
\be
c_L
\; = \; 1 + 6 \left( b + \frac{1}{b} \right)^2
\; = \; 13 + 6 \left( b^2 + b^{-2} \right)
\label{cLVir}
\ee
where the screening parameters $(b^2,b^{-2})$ are sometimes denoted $(\alpha,\alpha^{-1})$.
Note, that $\text{Vir}_b$ is an example of a {\it vertex algebra with parameters}: a family of VOAs, such that each member of the family has the same set of fields, but structure constants that depend on a parameters (in this case, parameter $b$). In particular, the central charge of such VOAs may also depend on the parameters, much like \eqref{cLVir}.
Then, essentially by the BPS/CFT and AGT correspondence \cite{Nikitatalk,Nekrasov:2015wsu,Alday:2009aq},
we have $\text{Vir}_b = \text{VOA} [M_4]$ when $M_4$
is a Euclidean space with the $\Omega$-background \cite{Nekrasov:2002qd}:
\be
\R^4_{\epsilon_1,\epsilon_2}
\quad \simeq \quad B^4
\quad \simeq \quad ~{\raisebox{-.6cm}{\includegraphics[width=2.6cm]{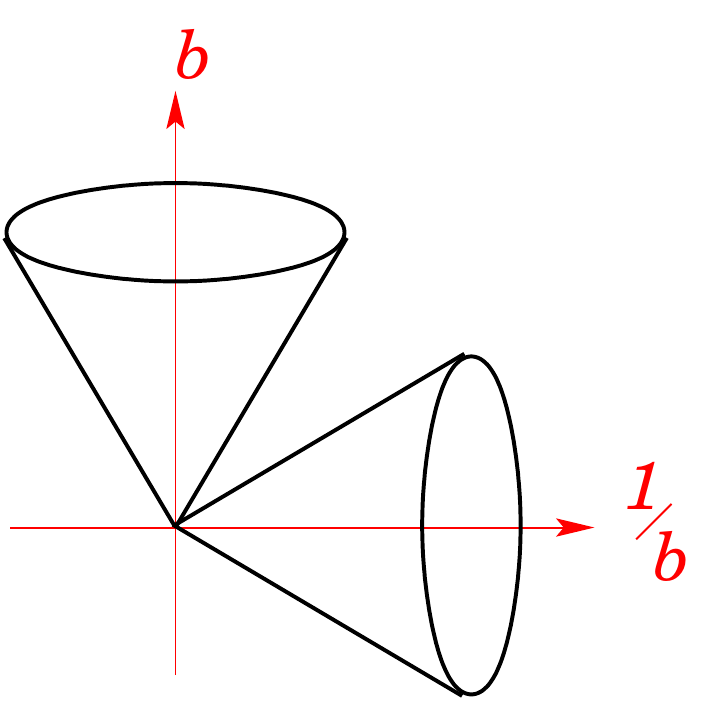}}}
\label{flatpatch}
\ee
It can be used as a patch (a 4-ball) to build more general 4-manifolds.
Note, however, that unlike other parts of this paper, where $M_4$ simply stands for smooth 4-manifold, the $\Omega$-background discussed here is not part of the smooth structure data. Rather, it is an extra structure associated with $U(1) \times U(1)$ torus action on $\R^4 \simeq B^4$, which is the key element in the physical formulation of the $\Omega$-background~\cite{Nekrasov:2002qd}. (See \cite{MaulikO,SVasserot} for a mathematical formulation of the problem). In other words, we work with $\R^4 \simeq B^4$ equivariantly.
This is what notation $\R^4_{\epsilon_1,\epsilon_2}$ in \eqref{flatpatch} indicates, with $\epsilon_1$ and $\epsilon_2$ being the equivariant parameters for $U(1) \times U(1)$ torus action which, too, is often denoted in a similar way by $U(1)_{\epsilon_1} \times U(1)_{\epsilon_2}$.
Moreover, below we require that this structure is preserved by gluing and, for this reason, in the rest of this section $M_4$ needs to be further assumed to be a toric 4-manifold.

As a toric manifold, $\R^4_{\epsilon_1,\epsilon_2} \cong \C^2_{\epsilon_1,\epsilon_2}$ with its natural $U(1)_{\epsilon_1} \times U(1)_{\epsilon_2}$ action has two complex lines, $\R^2_{\epsilon_i} \cong \C_{\epsilon_i}$, fixed by each of the two $U(1)$ factors in $U(1)_{\epsilon_1} \times U(1)_{\epsilon_2}$.
These two lines (that we sometimes call ``legs'') are illustrated in the right-hand side of \eqref{flatpatch}.
On the VOA side, these two ``legs'' labeled by $b$ and $1/b$ correspond to the two types of primary operators $\phi_{n,1}$ and $\phi_{1,n}$
which can be used to form non-trivial extensions of the tensor products of Virasoro algebras, {\it cf.} \cite{Bershtein:2013oka}.
Therefore, it is convenient to introduce the following graphical representation
for $\text{Vir}_b = \text{VOA} [\R^4_{\epsilon_1,\epsilon_2}]$:
\be
\text{Vir}_b \quad \simeq \quad ~{\raisebox{-.4cm}{\includegraphics[width=1.6cm]{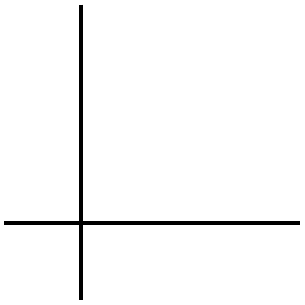}}}
\label{Virpatch}
\ee

\begin{figure}[ht]
	\centering
	\includegraphics[width=1.2in]{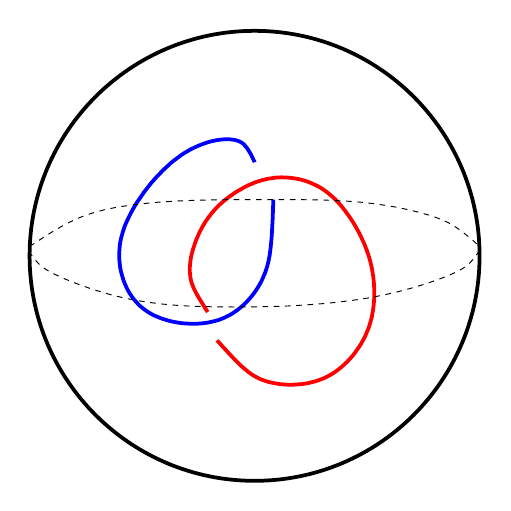}
	\caption{An illustration of a Hopf link on the $S^3$ boundary of a 4-ball.}
	\label{fig:Hopf}
\end{figure}

For purposes of gluing \eqref{flatpatch} into more general (toric) 4-manifolds, one should think of it as a 4-ball $B^4$. Of course, it still has to be considered equivariantly, {\it i.e.} as a toric manifold with $U(1)_{\epsilon_1} \times U(1)_{\epsilon_2}$ action.
The two lines $\R^2_{\epsilon_i} \cong \C_{\epsilon_i}$ fixed by the $U(1)$ factors and illustrated in the graphical notations \eqref{flatpatch}--\eqref{Virpatch} meet the boundary $S^3 = \partial B^4$ along the Hopf link, as shown in Figure~\ref{fig:Hopf}.
And, since we require gluing to preserve the toric structure, attaching such 4-balls must be done along the components of the Hopf link.
In the language of Kirby calculus~\cite{GompfS}, the circles represented by the components of the Hopf link in Figure~\ref{fig:Hopf} are the ``attaching circles'' of the 2-handles in the construction of toric 4-manifold $M_4$ from local patches \eqref{flatpatch}.

\bigskip

On the algebra side, gluing two copies of \eqref{Virpatch} together corresponds to extending $\text{Vir}_{b_1} \otimes \text{Vir}_{b_2}$
by operators of the form $\phi_{1,n} \otimes \phi_{n,1}$, {\it etc.}:
\be
~{\raisebox{-.4cm}{\includegraphics[width=1.6cm]{Lshape}}}
\!\!\!\!\!\!\!\!~{\raisebox{-.4cm}{$\alpha_1$}}
\quad\quad
~{\raisebox{-.4cm}{$\alpha_2^{-1}$}}\!\!\!\!\!\!\!\!\!\!\!\!
~{\raisebox{-.4cm}{\includegraphics[width=1.6cm]{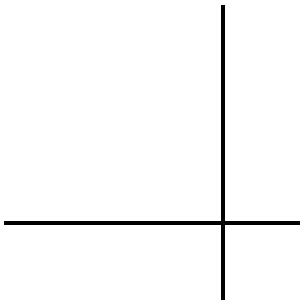}}}
\label{VirVirgluing}
\ee
Requiring absence of monodromy leads to the condition
\be
\alpha_1 + \alpha_2^{-1} \; \in \; \Z
\label{nomonodromy}
\ee
which then becomes the basic gluing condition.
A topological interpretation of the integer on the right-hand side will be given shortly.

\begin{example}{$M_4 = \cp^2 \setminus \{ \text{pt} \}$}
As a simple example of \eqref{nomonodromy}, consider
$\alpha_1 + \alpha_2^{-1} = b_1^2 + b_2^{-2} = +1$.
Using \eqref{cLVir} and writing solutions to this constraint as $b_1 = \frac{b}{\sqrt{b^2-1}}$ and $b_2 = \sqrt{1-b^2}$,
we obtain the following expression for the central charge
\begin{eqnarray}
c_L \left( \, \text{Vir}_{b_1} \otimes \text{Vir}_{b_2} \, \right)
& = & 13 + 6 \left( b_1^2 + b_1^{-2} \right)
+ 13 + 6 \left( b_2^2 + b_2^{-2} \right) \label{cLalmostCP2} \\
& = & 44 - \frac{6}{b^2} - 6 b^2
\; = \; 57 - \left[ 1 + 6 \left( b + \frac{1}{b} \right)^2 \right]
\nonumber
\end{eqnarray}
%\begin{multline}
%c_L \left( \, \text{Vir}_{b_1} \otimes \text{Vir}_{b_2} \, \right)
%\; = \; 13 + 6 \left( b_1^2 + b_1^{-2} \right)
%+ 13 + 6 \left( b_2^2 + b_2^{-2} \right) \\
%\; = \; 44 - \frac{6}{b^2} - 6 b^2
%\; = \; 57 - \left[ 1 + 6 \left( b + \frac{1}{b} \right)^2 \right]
%\label{cLalmostCP2}
%\end{multline}
which according to our gluing rules corresponds to
\be
M_4 \quad \simeq \quad\quad ~{\raisebox{-.4cm}{\includegraphics[width=1.8cm]{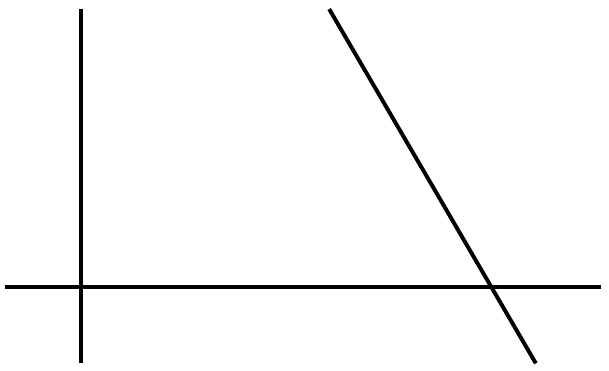}}}
\label{MalmostCP2}
\ee
Indeed, it is easy to see that adding another patch \eqref{flatpatch} completes \eqref{MalmostCP2}
into a toric diagram of $\cp^2$, and the central charge \eqref{cLalmostCP2} to the corresponding value $c_L = 57$
listed in Table~\ref{table:cLcR}.
\end{example}

\subsection{Plumbing graphs}

The simple examples above admit a generalization to a large class of 4-manifolds --- and, correspondingly, vertex algebras ---
labeled by graphs, whose vertices are decorated by integer numbers.
%(These integers are called Euler numbers or framing coefficients.)

Basically, vertices of the graph are in one-to-one correspondence with the generators $S_i$ of $H_2 (M_4,\Z)$.
In particular, the total number of vertices is equal to $b_2 (M_4)$.
Integer labels of the vertices, called Euler numbers or framing coefficients, are self-intersection numbers $a_i := S_i \cdot S_i$,
whereas each pair of vertices $i$ and $j$ connected by an edge means that $S_i$ and $S_j$ intersect, $S_i \cdot S_j = 1$.

A more accurate description of a 4-manifold associated to a plumbing graph is obtained by replacing the latter with
a link in $S^3$, as illustrated in Figure~\ref{fig:Anplumbing}, and interpreting the result as a Kirby diagram of a 4-manifold.
In other words, each vertex with integer label $a_i$ corresponds to a link component ($\cong$ unknot)
with framing $a_i$ which, in turn, represents the attaching circle of a 2-handle with Euler number $a_i$.

In order to see how such data defines a vertex algebra,
suppose we perform the gluing \eqref{VirVirgluing} a total of $n$ times,
each time meeting the compatibility condition \eqref{nomonodromy}:
\be
\alpha_i + \frac{1}{\alpha_{i+1}} \; = \; a_i
\qquad\qquad a_i \; \in \; \Z
\label{aaasum}
\ee
The result of this process will be a vertex algebra with parameters $\frac{1}{\alpha_1}$ and $\alpha_{n+1}$
related by the continued fraction:
\be
\alpha_1 \; = \; a_{1} - \cfrac{1}{a_2 - \cfrac{1}{\ddots - \cfrac{1}{\alpha_{n+1}}}}
\label{aacontfrac}
\ee
What is the corresponding 4-manifold $M_4$?

We already know the answer to this question at least in two special cases, when $n=1$ and $a_1 = \pm 1$.
The choice $a_1 = +1$ is precisely the above example \eqref{MalmostCP2}, where $M_4$ is a disk bundle over $S^2$ with Euler number $+1$.
On the other hand, for $a_1 = -1$ it was argued in \cite{Bershtein:2013oka} that the corresponding algebra --- called ``Urod" algebra $\CU$ ---
corresponds to a 4-manifold $M_4 = \bar \cp^2 \setminus \{ \text{pt} \}$, {\it i.e.} a disk bundle over $S^2$ with Euler number $-1$.
In each case, the corresponding 4-manifold is a disk bundle over $S^2$ with Euler number $a_1 = \pm 1$.
Therefore, more generally, that is for any $a_i \in \Z$,
one might expect that an algebra built this way corresponds to a disk bundle over $S^2$
with Euler number $a_1$ or, if $n>1$, to a plumbing of 2-spheres with Euler numbers $a_i$, $i = 1, \ldots, n$:
\begin{figure}[ht]
\centering
\includegraphics[width=4.5in]{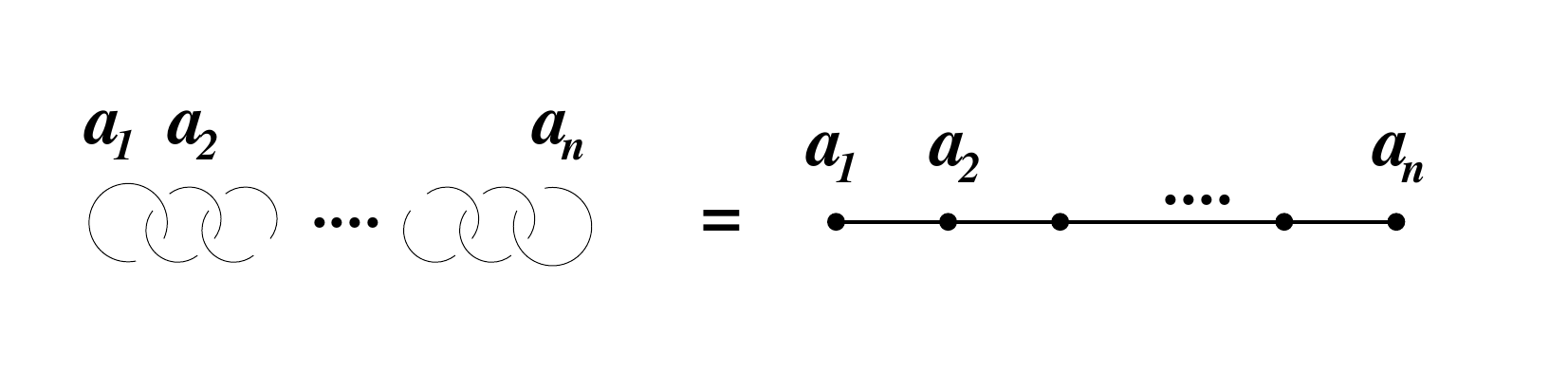}
\caption{A Kirby diagram and the corresponding plumbing graph.}
\label{fig:Anplumbing}
\end{figure}

Indeed, a low-dimensional topologist would immediately associate the continued fraction \eqref{aacontfrac}
to the 4-manifold whose Kirby diagram is shown in Figure~\ref{fig:Anplumbing}, even before looking at the special cases.
Yet, we wish to subject our proposal to further tests.
To that end, let us consider a more non-trivial version of the gluing operation \eqref{VirVirgluing} with
\be
\alpha_1 + \alpha_2^{-1} \; = \; -2
\ee
According to our proposal, we expect to find a VOA associated to a 4-manifold $M_4 \cong T^* S^2$ or,
equivalently, the total space of $\CO (-2)$ bundle over $\cp^1$.
In algebraic geometry, this space is known as the resolution of $A_1$ singularity
and in what follows we also refer to it either using a Kirby diagram or the corresponding plumbing graph:
\be
A_1
\qquad = \qquad
{-2 \atop \unknot}
\qquad = \qquad
{-2 \atop \bullet}
\ee
The corresponding algebra is
\be
\text{VOA} \Big[ \; {-2 \atop \unknot} \; \Big]
~~=~~
\text{toroidal Yangian of }\frak{gl}_2
\label{A1toroidalY}
\ee
Indeed, much like the ordinary Yangian \cite{Olshanski},
this algebra can be defined as analytic continuation with respect to $k$
of the coset of the affine Lie algebra $\hat{\frak{gl}}_k$ by its subalgebra
$\hat{\frak{gl}}_{k-2} \subset \hat{\frak{gl}}_k$, both at level $N \in \C$.
We denote this algebra by $\CC_N (\hat{\frak{gl}}_{k}, \hat{\frak{gl}}_{k-2})$.
(Note, the role of $N$ and $k$ here is exchanged compared to \cite{Feigin:2013fga}, where a similar notation was used.)
Clearly, this coset algebra can be described as a conformal extension of two simpler (``one-step'')
cosets\footnote{Here, and in the next equation below, ``$\boxtimes$'' is used to denote extension,
which usually does not have a designated symbol since, after all,
extension is not a unique operation (there can be many different extensions).
In the case at hand, however, there is no confusion and it is clear which extension of the cosets
$\CC_N (\hat{\frak{gl}}_{k}, \hat{\frak{gl}}_{k-1})$ and $\CC_N (\hat{\frak{gl}}_{k-1}, \hat{\frak{gl}}_{k-2})$
gives $\CC_N (\hat{\frak{gl}}_{k}, \hat{\frak{gl}}_{k-2})$.}
\be
\CC_N (\hat{\frak{gl}}_{k}, \hat{\frak{gl}}_{k-2})
\; = \;
\CC_N (\hat{\frak{gl}}_{k}, \hat{\frak{gl}}_{k-1})
\; \boxtimes \;
\CC_N (\hat{\frak{gl}}_{k-1}, \hat{\frak{gl}}_{k-2})
\label{Yangcosets}
\ee

On the other hand, according to \eqref{VirVirgluing},
the algebra that we associate to gluing two copies of \eqref{flatpatch} is an extension
of two Virasoro algebras or, more generally,
for $G = U(N)$,
two $W$-algebras\footnote{Recall, that $W$-algebra $\CW ({\frak g})$ can be defined as a result of the quantum Drinfeld-Sokolov reduction of the $\hat {\frak g}$ current algebra (WZW model in the physics literature). See {\it e.g.} \cite{FF} for details. In the special case ${\frak g} = \frak{gl}_2$, it gives $\CW (\frak{gl}_2) = \text{Vir} \oplus \text{Heisenberg}$.}
$\CW ({\frak g})$ with ${\frak g} = \frak{gl}_N$:
\be
\text{VOA} \Big[ \; {-2 \atop \unknot} \; \Big]
\; = \;
\CW (\frak{gl}_N)
\; \boxtimes \;
\CW (\frak{gl}_N)
\label{O2WW}
\ee
This is in perfect agreement with \eqref{Yangcosets} due to a well-known isomorphism\footnote{There are several ways to define the $W$-algebra. One is the quantum Drinfeld-Sokolov reduction of $\hat{\frak{gl}}_{n}$ at level $k$. It gives the algebra that depends on $n$ and $k$, in which one can make an analytic continuation with respect to $n$. Another convenient parametrization can be done in terms of three parameters $h_1$, $h_2$, $h_3$ that add up to zero. (These are Nekrasov's equivariant parameters.) Another parameter is $h$ (the ``level''). They all are related as $h_1/h_2= 1/(n+k)$, $-h/h_3 =n$. One can define the $W$-algebra for superalgebra $\hat{\frak{gl}} (n|m)$. Then, if $k_1$ and $k_2$ are the levels of $\hat{\frak{gl}} (n)$ and $\hat{\frak{gl}} (m)$, respectively, we have $\frac{1}{n+k_1} +\frac{1}{m+k_2}=1$ and the DS reduction gives $\hat{\frak{gl}} (n)$ type algebra with parameters $h_1/h_2=1/(n+k_1)$, $h_3/h_2=1/(m+k_2)$ and $h= - (n h_3 + m h_1)$. Finally, as discussed in the main text, the coset $\hat{\frak{gl}} (m) / \hat{\frak{gl}} (m-1)$ at level $n$ gives the $W$-algebra of type $\hat{\frak{gl}} (n)$ at level $k$, such that $\frac{m+n}{m+n-1} = \frac{1}{k+n}$; this also happens to be $\hat{\frak{gl}} (n|n-1)$ $W$-algebra. See {\it e.g.} \cite{Feigin:2013fga} for many of the relations quoted here.} \cite{Feigin:2013fga}:
\be
\CC_N (\hat{\frak{gl}}_{k}, \hat{\frak{gl}}_{k-1})
\; \simeq \;
\CW_{N,\frac{N+k+1}{N+k+2}}
\ee
where $\CW_{N,\frac{1}{N+\kappa}}$ denotes the result of the quantum Drinfeld-Sokolov reduction
of $\hat{\frak{gl}}_{N}$ at level $\kappa$, analytically continued with respect to $N$ \cite{FF}.
Note, under this isomorphism, the rank of the ``gauge group'' $G = U(N)$ becomes the level
of the coset construction $\CC_N (\hat{\frak{gl}}_{k}, \hat{\frak{gl}}_{k-2})$,
whereas $k$ is related to the equivariant parameters of the $U(1)_{\epsilon_1} \times U(1)_{\epsilon_2}$ action on $M_4$.

We conjecture that, for more general $\CO(-p)$ bundles over $\cp^1$,
the analogue of \eqref{A1toroidalY} is given by the shifted affine Yangian of $\frak{gl}_2$.

\bigskip
\begin{figure}[ht]
\centering
\includegraphics[width=1.7in]{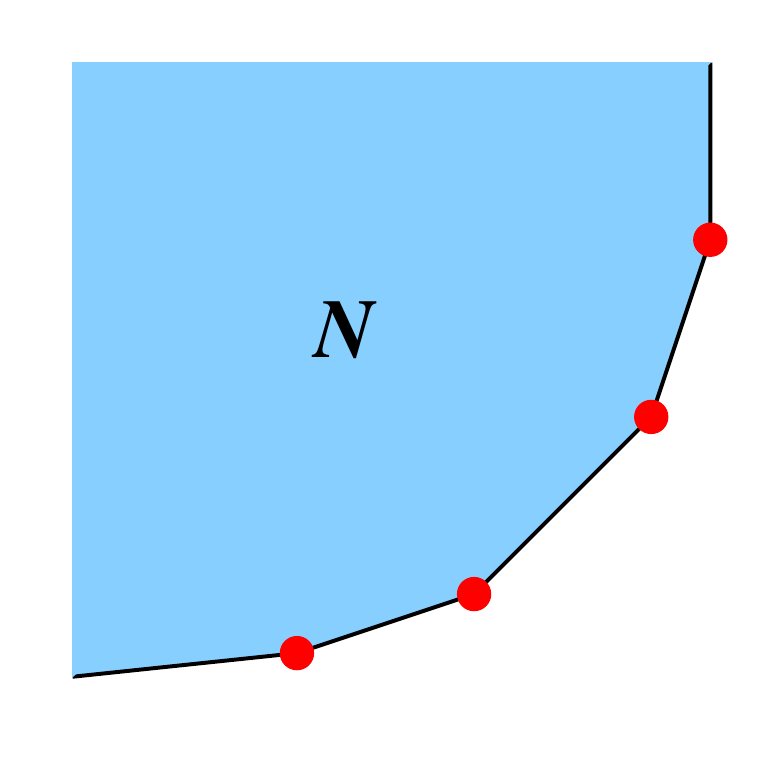}
\caption{Toric diagram for a linear plumbing.}
\label{fig:Antoric}
\end{figure}

A continued fraction \eqref{aacontfrac} or, equivalently, a linear plumbing graph
determine a sequence of integer pairs $(p_i,q_i)$, $i=0,\ldots, n+1$,
that can be conveniently visualized
as lattice vectors $\vec v_i = (p_i,q_i)$ on a two-dimensional plane.
These lattice vectors form the boundary of a toric diagram of $M_4$,
illustrated in Figure~\ref{fig:Antoric}, and can be determined recursively:
\be
\vec v_{i+1} + a_i \vec v_i + \vec v_{i-1} \; = \; 0
\qquad \qquad
1 \le i \le n
\ee
A slightly more explicit relation in terms of continued fractions reads ($1 \le i \le n$):
$$
- \frac{q_{i-1}}{q_i} \; = \;
a_{i} - \cfrac{1}{a_{i+1} - \cfrac{1}{\ddots - \cfrac{1}{a_{n}}}}
\qquad,\qquad
- \frac{p_{i+1}}{p_i} \; = \;
a_{i} - \cfrac{1}{a_{i-1} - \cfrac{1}{\ddots - \cfrac{1}{a_{1}}}}
$$
with $p_0 = 0$, $p_1 = 1$, $q_n = 1$, and $q_{n+1}=0$.
For example, in the basic case $n=1$, from these recursive relations we learn that
\be
{a_1 \atop \unknot}
\quad: \qquad
\left\{
\begin{aligned}
(p_0,q_0) & = (0,-a_1) \\
(p_1,q_1) & =(1,1) \\
(p_2,q_2) & =(-a_1,0)
\end{aligned}
\right.
\ee
This description of toric $M_4$ in terms of lattice vectors $(p_i,q_i)$
is useful not only for building $\text{VOA} [M_4]$ via ``gluing,''
but also for understanding various dual brane configurations in string theory.

Indeed, the defining property of a toric 4-manifold $M_4$ is that it is fibered by 2-dimensional tori,
which degenerate along the skeleton of a toric diagram.
And, one of the well known string dualities asserts that M-theory on a 2-torus $T^2$
is equivalent to type IIB string theory on a circle $S^1$, whose radius is inversely
proportional to the volume of the M-theory torus.
This duality can be equivalently described as a composition of
a simpler relation between type IIA string theory and M-theory on $S^1$
and the standard T-duality between type IIA and type IIB string theories.
Whichever description we use, it is easy to see that $N$ fivebranes wrapped
on the 2-torus in M-theory map to $N$ D3-branes in type IIB string theory.

Moreover, the degeneration locus of the $T^2$ fiber, {\it i.e.} the skeleton of
the toric diagram, maps to a web of $(p,q)$ fivebranes in type IIB string theory that
carry $p$ units of the NS5-brane charge and $q$ units of the D5-brane charge.
Therefore, for toric $M_4$ our original physical setup \eqref{TM4def} is dual
to a configuration of $N$ D3-branes,
with boundary on a network of $(p_i,q_i)$ 5-branes in type IIB string theory:
\be
\begin{array}{rcl}
N \; \text{fivebranes} \; & \; \Rightarrow \; & \; N \; \text{D3-branes} \vspace{.05in} \\
\text{toric diagram} \; & \; \Rightarrow \; & \; \text{web of} \; (p_i,q_i) \; \text{5-branes}
\end{array}
\ee

Therefore, our proposal for gluing plumbing graphs, at least in the toric case
can be equivalently formulated as a statement about gluing vertex algebras
at the ``corners'' of fivebrane web in type IIB string theory.
In the special case, when $p_i q_{i+1} - p_{i+1} q_i = 1$,
a proposal for such vertex algebras was recently made in \cite{BFMerzon,Litvinov:2016mgi,Gaiotto:2017euk,Prochazka:2017qum},
and
$\text{VOA} \big[
\overset{\displaystyle{a_1}}{\bullet}
\frac{\phantom{xx}}{\phantom{xx}}
\overset{\displaystyle{a_2}}{\bullet}
\frac{\phantom{x}}{\phantom{x}}
\cdots
\frac{\phantom{x}}{\phantom{x}}
\overset{\displaystyle{a_n}}{\bullet}
\big]$ can be viewed as a more general version of this proposal, for webs with arbitrary $p_i$ and $q_i$.
It would be interesting to check this proposal by studying the full physical 2d $\CN=(0,4)$
theory.
Note, the $\CN=(0,4)$ supersymmetry agrees with the fact that toric $M_4$ are K\"ahler,
which is when supersymmetry of the 2d super-conformal theory $T[M_4]$ is enhanced from $\CN=(0,2)$ to $\CN=(0,4)$. Our initial attempts\footnote{We thank Davide Gaiotto for useful discussions on this point.} to promote VOAs in \cite{Gaiotto:2017euk,Prochazka:2017qum}, and their generalizations in \cite{Creutzig:2017uxh,Frenkel:2018dej}, to physical 2d $\CN=(0,4)$ superconformal theories turned out to be more challenging than one might have expected based on the duality between brane systems in M-theory and type IIB string theory described in the above.
This is especially surprising since trialities, similar to those of \cite{Gaberdiel:2012uj,Gaiotto:2017euk}, can be realized in two-dimensional SCFTs with $\CN=(0,2)$ and $\CN=(0,1)$ supersymmetry \cite{Gadde:2013lxa,trialities}; the former are directly relevant to $T[M_4]$ and will be discussed further below in section~\ref{sec:3manifolds}.
With these motivations in mind, it would be good to mobilize the search for trialities in 2d SCFTs with $\CN=(0,4)$ supersymmetry.

The class of models discussed here admits a generalization in which
$T^2$ fiber of a toric $M_4$ is replaced by a surface $F_g$ of arbitrary genus $g$.
Surprisingly, this generalization allows to describe an arbitrary smooth 4-manifold.
This will be the subject of section~\ref{sec:trisections}
where we also explain how the chiral algebra $\text{VOA}[M_4]$
can be constructed from basic ingredients that replace brane webs:
namely, these will be junctions of Heegaard boundary conditions (``Heegaard branes'').

\subsection{Toric manifolds and toroidal algebras}

Starting in \eqref{VOAmoduli} we set out to explore algebraic structures acting on (generalized) cohomology
of moduli spaces of $G$-instantons on $M_4$.

When $M_4$ is a toric surface, the general gluing construction, illustrated in \eqref{VirVirgluing} or \eqref{O2WW},
leads to various extensions of $W$-algebras for $\frak{g}_{\C} = \text{Lie} (G_{\C})$ of the same type as the original ``gauge'' group $G$.
Although this general gluing construction is supported by a number of direct
calculations \cite{Bonelli:2012ny,Bruzzo:2013daa,Bershtein:2016mxz,Bawane:2014uka},
which build on the AGT conjecture \cite{Alday:2009aq} for $M_4 = \R^4_{\epsilon_1,\epsilon_2}$, it may seem puzzling.
Indeed, when $G=U(N)$ and $M_4$ is a 4-manifold with a linear plumbing, as in Figure~\ref{fig:Anplumbing}:
\be
A_{n-1} \quad = \quad
\underbrace{
\begin{array}{cccc}
\overset{\displaystyle{-2}}{\bullet}
\frac{\phantom{xxxx}}{\phantom{xxxx}} &
\!\!\!\! \overset{\displaystyle{-2}}{\bullet}
\frac{\phantom{xxxx}}{\phantom{xxxx}}
& \cdots &
\frac{\phantom{xxxx}}{\phantom{xxxx}}
\overset{\displaystyle{-2}}{\bullet}
\end{array}
}_{n-1~\text{vertices}}
\label{Apmanifold}
\ee
the seminal work of Nakajima \cite{Nakajima} says that $\text{VOA}[M_4,G]$ contains $\hat{\frak{sl}}_{n}$ at level $N$.
These two algebraic structures appear to be completely different and almost in contradiction, unless there is a larger algebra that contains both.

In fact, we already saw a hint of such larger algebraic structure in \eqref{A1toroidalY},
where the rank of the ``gauge'' group $G$ turned into a level,
whereas topology of $M_4$ determined the type of the Lie algebra underlying $\text{VOA} [M_4, G]$.
Another hint comes from an observation that $\hat{\frak{sl}}_{n}$ at level $N$
does not depend on the equivariant parameters $\epsilon_1$ and $\epsilon_2$ --- in fact,
the equivariant $U(1)_{\epsilon_1} \times U(1)_{\epsilon_2}$ action on $M_4$
does not play much role in \cite{Nakajima} --- whereas extensions of $W$-algebras obviously do.

Therefore, it is probably not too surprising, after all, that different algebraic structures
answer slightly different questions, even if the original setup is similar.
Moreover, these hints suggests that a larger algebraic structure,
that contains both $\hat{\frak{sl}}_{n}$ at level $N$ and $W$-algebras of type $\frak{gl}_{N}$,
can be found in a combined equivariant setting.
Indeed, a natural candidate for a larger algebraic structure that has such properties is
the quantum toroidal algebra $\CE_{q_1,q_2,q_3} (\frak{g})$
(sometimes also denoted $\ddot \CU_{q_1,q_2,q_2} (\frak{g})$ or $\CU_{q_1,q_2} (\widehat{\widehat{\mathfrak{g}}})$),
for $\frak{g} = \frak{gl}_{N}$ in the context of the present discussion.

For a given simple Lie algebra $\frak{g}$, the quantum toroidal algebra $\CE_{q_1,q_2,q_3} (\frak{g})$
is a quantization \cite{GKV} of the toroidal Lie algebra (= two-dimensional central extension
of the double loop algebra ${\frak g} \otimes [z^{\pm 1}, w^{\pm 1}]$),
much like the ordinary quantum affine algebra $\CU_q (\hat{\frak g})$
is a quantization of the affine Kac-Moody algebra $\hat {\frak g}$.
It depends on three complex parameters, which satisfy
\be
q_1 , q_2, q_3 \; \in \; \C^*
\qquad\qquad
q_1 q_2 q_3 \; = \; 1
\label{q1q2q3}
\ee
and in general has two central elements (denoted $q^c$ and $\kappa$ in \cite{Feigin:2013fga}).
Moreover, for $N > 1$, the quantum toroidal $\frak{gl}_{N}$ is symmetric in $q_1$ and $q_3$ (but not $q_2$).
A particular way to solve \eqref{q1q2q3}
\be
q_1 = \frac{d}{q} \,, \qquad
q_2 = q^2 \,, \qquad
q_3 = \frac{1}{qd}
\ee
leads to another notation $\ddot \CU_{q,d} (\frak{g})$ for the quantum toroidal algebra.

A remarkable feature of the quantum toroidal algebra is that it has an automorphism $S$ which, among other things,
exchanges the role ``horizontal'' and ``vertical'' generators, see {\it e.g.} \cite{FJMM,Feigin:2017wnq}.
In particular, $\CE_{q_1,q_2,q_3} (\frak{g})$ contains {\it two} copies
of the quantum affine algebra $\CU_q (\hat{\frak g})$, often called ``horizontal'' and ``vertical,''
\be
h,v : \quad \CU_q (\hat{\frak{g}}) \longrightarrow \CE_{q_1,q_2,q_3} (\frak{g})
\label{Sverthoriz}
\ee
which are exchanged by the automorphism $S$.

The toroidal Yangian (a.k.a. affine Yangian) that we encountered in \eqref{A1toroidalY}
is the ``additivization'' of $\CE_{q_1,q_2,q_3} (\frak{g})$, obtained from the latter
in the {\it conformal limit}:
\be
q_1 = \epsilon^{h_1}
\,,\quad
q_2 = \epsilon^{h_2}
\,,\quad
q_3 = \epsilon^{h_3}
\,,\quad
\kappa = \epsilon^k
\,,\quad
\epsilon \to 1
\label{conflimitY}
\ee
The resulting Yangian, usually denoted $\hat Y_{h_1,h_2,h_3} (\frak{g})$ or $\ddot Y_{h_1,h_2,h_3} (\frak{g})$,
is a two-parameter deformation of the universal enveloping algebra
of the universal central extension of $\frak{sl}_n [s^{\pm 1},t]$,
much like the ordinary Yangian $Y_{\hbar} (\frak{g})$, introduced by Drinfeld,
is a deformation of the universal enveloping algebra $\CU (\frak{g} [z])$.
In particular, $\hat Y_{h_1,h_2,h_3} (\frak{g})$ has parameters, {\it cf.} \eqref{q1q2q3}:
\be
h_1 , h_2, h_3 \; \in \; \C
\qquad\qquad
h_1 + h_2 + h_3 \; = \; 0
\ee
We can summarize the relation between $\CE_{q_1,q_2,q_3} (\frak{g})$ and $\hat Y_{h_1,h_2,h_3} (\frak{g})$
in the following diagram\footnote{Another way to obtain $\hat Y_{h_1,h_2,h_3} (\frak{sl}_n)$ as
a limit of the quantum toroidal algebra $\CE_{q_1,q_2,q_3} (\frak{gl}_1)$
was proposed in \cite{Belavin:2012eg}, based on $q_1 \to e^{2\pi i /n}$ and $q_2 \to 1$.}
\be
\begin{CD}
\CE_{q_1,q_2,q_3} (\frak{g})   @>\text{~~~~conformal limit~~~~}>>   \hat Y_{h_1,h_2,h_3} (\frak{g}) \\
@A{\text{quantization}}AA                                    @AA{\text{quantization}}A\\
\CU (\frak{g} [z^{\pm 1}, w^{\pm 1}])                  @.      \CU (\frak{g} [z^{\pm 1}, w])
\end{CD}
\ee

To summarize, the toroidal Yangian $\hat Y_{h_1,h_2,h_3}$ has two rather different descriptions: one starts with the affine Kac-Moody algebra $\hat {\frak g} = \hat{\frak{sl}}_{n}$, whereas the other involves extensions of rank-$N$ $W$-algebras. Moreover, since the roles of the level and rank is interchanged in these two descriptions, we can say that they are related by ``level-rank duality.'' Although not all algebraic structures of the toroidal Yangian ({\it e.g.} comultiplication) are manifest in the coset description used in \eqref{Yangcosets},
\be
\hat Y_{h_1,h_2,h_3} (\frak{gl}_n) \; \cong \; \CC_N (\hat{\frak{gl}}_{k}, \hat{\frak{gl}}_{k-n})
\label{Ytorcos}
\ee
it still provides one of the most convenient descriptions where several algebraic structures relevant to us here can be seen simultaneously, see \cite{Feigin:2013fga,FJM} for details.
{}From the viewpoint of the physical system \eqref{TM4def}, these two different descriptions of \eqref{Ytorcos} arise as two different limits:
in the ``zero radius'' (orbifold) limit of \eqref{Apmanifold} the physical system has manifest
$SU(n)$ symmetry, which is enhanced to the affine algebra $\hat{\frak{sl}}_{n}$ in the 2d theory $T[M_4]$.
On the other hand, the ``large volume'' limit of \eqref{Apmanifold} leads to extensions of $W$-algebras.
Furthermore, the symmetry between the horizontal and vertical generators \eqref{Sverthoriz},
that plays an important role here,
becomes manifest only at the level of the quantum toroidal algebra, before taking the conformal limit \eqref{conflimitY}.
(See also \cite{Awata:2017lqa} and references therein.)

%%%%%%%%%%%%%%%%%%%%%%%%%%%%%%%%%%%%%%%%%%%%%%%%%%%%%%%%%%%%%%%%%%%%%%%%%%%%%%%%%%%%

\begin{figure}[ht]
\centering
\includegraphics[width=3.7in]{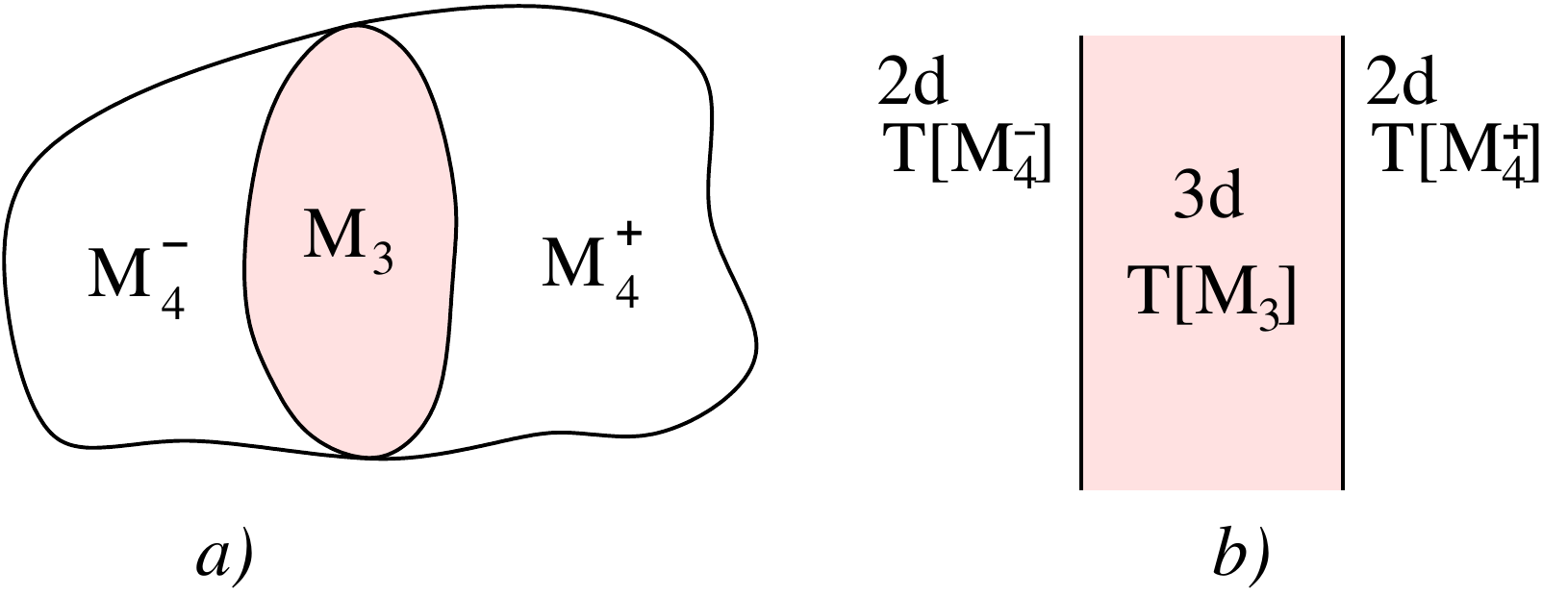}
\caption{$(a)$ Two 4-manifolds glued along a common boundary $M_3 = \pm \partial M_4^{\pm}$ correspond to $(b)$
three-dimensional $\CN=2$ theory $T[M_3]$ sandwiched by the 2d $\CN=(0,2)$ boundary conditions $T[M_4^-]$ and $T[M_4^+]$.}
\label{fig:M4d2d}
\end{figure}

\section{Operations and relations}
%\section{Gluing along 3-manifolds}
\label{sec:3manifolds}

\subsection{Generalized blow-ups}

Gluing along $M_3 = S^3$ basically corresponds to taking a tensor product of the VOAs associated with 4-manifold pieces that are being glued.
More precisely, the operation of taking the connected sum of two 4-manifolds involves removing a small 4-ball $B^4$ from both pieces and constructing a new 4-manifold by identifying their $S^3$ boundaries.
The simplest example of such operation is a (generalized) blow-up, which involves $\bar \cp^2 \setminus B^4$ or $\cp^2 \setminus B^4$ as one of the pieces and describes the connected sum with $\bar \cp^2$ or $\cp^2$.
Denoting $\text{VOA} [\bar \cp^2 \setminus B^4]$ and $\text{VOA} [\cp^2 \setminus B^4]$ by $\CU$ and $\bar \CU$, respectively, we can describe these blow-up operations as
\be
\text{VOA} \Big[ \; M_4 \, \# \, \bar \cp^2 \; \Big]
~~\cong~~
\CU \otimes \text{VOA} [M_4]
\label{blowupU}
\ee
and, similarly,
\be
\text{VOA} \Big[ \; M_4 \, \# \, \cp^2 \; \Big]
~~\cong~~
\bar \CU \otimes \text{VOA} [M_4]
\label{blowupUbar}
\ee
Note, that on the right-hand side of these relations we have $\text{VOA} [M_4]$ and not $\text{VOA} [M_4 \setminus B^4]$.
The central charges of vertex algebras $\CU$ and $\bar \CU$ can be easily deduced from the data already listed in Table~\ref{table:cLcR}:
\be
c_L (\CU) \; = \; -5
\qquad , \qquad
c_L (\bar \CU) \; = \; +31
\label{ccUrodUbar}
\ee
In fact, $\CU$ is precisely the ``Urod'' algebra \cite{Bershtein:2013oka} relevant to the blow-up of $M_4$,
whereas $\bar \CU$ is its close cousin with respect to exchanging the role of $b_2^+ (M_4)$ and $b_2^- (M_4)$.
We already encountered one description of the algebra $\bar \CU$ as extension of two copies of the Virasoro algebra
in example \eqref{MalmostCP2} discussed earlier.
The algebra $\bar \CU$ can be constructed in the same way as $\CU$ was constructed in \cite{Bershtein:2013oka}.
Note, even though such constructions involve VOAs with parameters as building blocks, the dependence on these parameters drops out in the resulting algebras $\CU$ and $\bar \CU$, as illustrated {\it e.g.} by their central charges \eqref{ccUrodUbar}.
In particular, as explained in \cite{Bershtein:2013oka}, any value of $b$ can be used in such extension constructions.

Embedding the construction of blow-up algebras $\CU$ and $\bar \CU$ into physical framework \eqref{TM4def}
suggests several interesting generalizations.
In one direction, it suggests the existence of higher-rank analogues of $\CU$ and $\bar \CU$, whose properties can be
inferred from \eqref{TM4def}.
For example, it predicts the values of the central charge $c_L (\CU) = - N^3 + 2N -1$ for $G = SU(N)$:
\be
\begin{array}{c|c|c|c|c|c}
G & SU(2) & SU(3) & SU(4) & SU(5) & \ldots \\
\hline
~c_L (\CU)~ & ~-5~ & -22~ & ~-57~ & ~-116~ & ~\ldots~
\end{array}
\ee
and, similarly, $c_L (\bar \CU) = 5N^3 - 4N - 1$.

A generalization in a different direction, that also follows from \eqref{TM4def}, is a prediction
that there exist supersymmetric CFTs which correspond to vertex algebras $\CU$ and $\bar \CU$,
and have central charges $c_R = -6$ and $c_R = 33$, respectively ({\it cf.} Table~\ref{table:cLcR}).
Indeed, as we already emphasized in the Introduction, all statements about chiral algebras $\text{VOA} [M_4]$
should have their counterparts in the 2d theory $T[M_4]$ which, in addition to the left-moving sector described by $\text{VOA} [M_4]$,
also has a right-moving $\CN=2$ supersymmetric sector.
This applies to all gluing operations and dualities (equivalences) discussed here,
including \eqref{blowupU} and \eqref{blowupUbar}.
It would be interesting to construct these supersymmetric CFTs explicitly.

We can also use \eqref{blowupU} in reverse to produce a natural candidate for $\text{VOA} [S^4]$ by combining it with the ``gluing rules'' of section~\ref{sec:extensions}.
Indeed, we can use the rules of section~\ref{sec:extensions} to first construct the $\text{VOA} [M_4]$ for $M_4 = \bar \cp^2$.
According to the general rules \eqref{nomonodromy}--\eqref{aaasum}, it is given by the extension of three Virasoro algebras, whose screening parameters $\alpha_i$, $i=1,2,3$, satisfy $\alpha_i + \alpha_i^{-1} = -1$. In other words, this example is similar to the one in \eqref{cLalmostCP2}--\eqref{MalmostCP2}, except that all Euler numbers $+1$ are now replaced by $-1$. Then, performing a blow-down operation, we obtain a description of the 4-sphere as a union of two 4-balls, $S^4 = B^4 \cup B^4$, glued along a common $S^3$ boundary in such a way that Euler numbers of the two lines (``legs'') are $a = -1 + 1 = 0$. (Note, that a blow-down operation removes one vertex from the ``toric diagram'' of $M_4 = \bar \cp^2$, which originally had a triangular shape, and changes the Euler numbres on the remaining two lines by $+1$.) Therefore, according to \eqref{nomonodromy}--\eqref{aaasum}, the corresponding VOA
is an extension of two Virasoro algebras, whose screening parameters $\alpha_1$ and $\alpha_2$
are related as
\be
\alpha_1 \; = \; - \frac{1}{\alpha_2}
\ee
It means that in the sum of the corresponding central charges \eqref{cLVir}
the $\alpha$-dependent terms neatly cancel, as in a typical system ``matter + gravity,'' and we obtain
\be
c_L \left( \text{VOA} [S^4] \right)
\; = \; 13 + 6 \left( \alpha_1 + \frac{1}{\alpha_1} \right) + 13 + 6 \left( \alpha_2 + \frac{1}{\alpha_2} \right)
\; = \; 26
\ee
in agreement with the value obtained by anomaly calculation in the string theory setup, {\it cf.} Table~\ref{table:cLcR}.

Physically, the 2d theory $T[S^4,G]$ and its left-moving chiral algebra $\text{VOA}[S^4,G]$
describe the ``center of mass'' degrees of freedom that should be present
(albeit possibly mixed with other degrees of freedom)
in $T[M_4,G]$ and $\text{VOA}[M_4,G]$ for any other closed 4-manifiold $M_4$.
After all, in a handle decomposition of $M_4$ one always finds a 0-handle and a 4-handle that make up a 4-sphere.
For $G=U(1)$, this center-of-mass multiplet is simply an ordinary 2d $\CN=(0,2)$ chiral multiplet with
\be
c_L \; = \; 2
\qquad , \qquad
c_R \; = \; 3 \,,
\label{cLcRfreechiral}
\ee
{\it i.e.} a complex $\C$-valued boson and a complex right-moving Weyl fermion \cite{Dedushenko:2017tdw}.
The complex boson is a coordinate on the fiber of $T^* \Sigma$ in \eqref{TM4def}.

For general $G$, all these fields become $\frak{g}$-valued, so that the resulting theory $T[S^4,G]$
is an interacting theory of a 2d $\CN=(0,2)$ adjoint chiral multiplet $\Phi$,
much like the analogous theory $T[S^3,G]$ in the context of 3d-3d correspondence~\cite{Gukov:2016gkn}.
For example, when $G=SU(2)$, there is still only one gauge-invariant $\CN=(0,2)$ superfield $\tr \Phi^2$,
but because the theory is no longer free the central charges are different from \eqref{cLcRfreechiral}
and both are shifted by $+24$ due to a $\CN=(0,2)$ Liouville type interaction.

\begin{figure}[ht]
\centering
\includegraphics[width=3.7in]{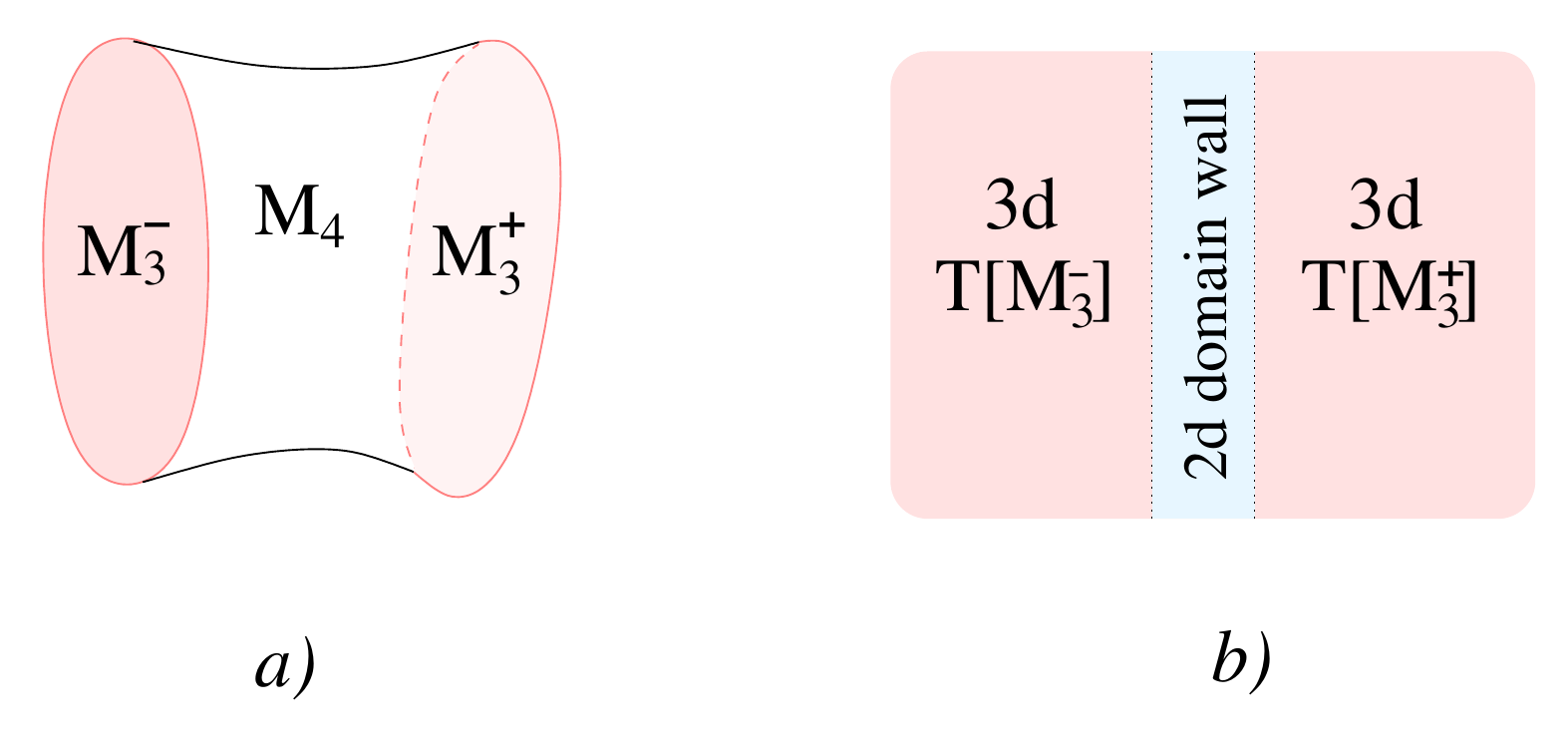}
\caption{$(a)$ A cobordism between 3-manifolds $M_3^-$ and $M_3^+$ corresponds to $(b)$
a 2d domain wall (interface) between 3d $\CN=2$ theories $T[M_3^-]$ and $T[M_3^+]$.}
\label{fig:M4d2dwall}
\end{figure}

In general, a 4-manifold with 3-dimensional boundaries $M_3^{(i)}$ defines an interface --- or, equivalently, via the folding trick ---
a boundary condition for the tensor product of 3d $\CN=2$ theories $T[M_3^{(i)}]$, see {\it e.g.} Figure~\ref{fig:M4d2dwall}
for an illustration in case of two boundary components.
Gluing such 4-manifolds with boundary in 3d $\CN=2$ theory corresponds to stacking the interfaces and / or sandwiching
boundary conditions, as illustrated in Figure~\ref{fig:M4d2d} for the simple case of two boundary conditions.
For a detailed discussion of the general case see \cite{Gadde:2013sca},
and for an application to yet another construction of $\text{VOA}[M_4]$ see section~\ref{sec:trisections}.

There are several slightly different version of gluing (reviewed in \cite{Gukov:2016gkn})
which differ by how one chooses to parametrize boundary conditions on $M_4$.
For example, there is a continuous basis, discrete basis, and many other choices.
In the correspondence between 4-manifolds and 2d $\CN=(0,2)$ theories,
these different choices show up as inclusion/exclusion of $T[M_3]$ degrees of freedom into $T[M_4]$.
Since $T[M_3]$ and $T[M_4]$ form a 2d-3d coupled system, there is no canonical choice.
Such choices play an important role in any relations that involve $M_4$ and its boundary,
{\it e.g.} \eqref{Krepring}, and in all gluing formulae where they affect the integration (convolution) measure.
Note, these different ways of parametrizing boundary conditions in QFT involve, among other things, group-theoretic data because the fields transform in representations of the gauge group $G$. For example, the so-called ``discrete basis'' mentioned above corresponds to parametrizing boundary conditions by the representations of the fundamental group \eqref{rhorep}. Other choices may have less natural geometric interpretation.

Of course, there is no such ambiguity for closed 4-manifolds without boundary;
aside from the obvious choice of ``gauge'' group $G$, the 2d conformal theory $T[M_4]$
and its chiral algebra $\text{VOA} [M_4]$ are unique for $b_2^+ > 1$.
Even when $b_2^+ = 1$ and there are wall crossing phenomena,
below we propose their interpretation in $\text{VOA} [M_4]$
and the physics of $T[M_4]$ is under good control.

Similarly, for a closed 3-manifold without boundary, the theory $T[M_3]$ is unique.
And, for gluing operations in Figures~\ref{fig:M4d2d} and \ref{fig:M4d2dwall},
it is crucial to use the correct (complete) theory $T[M_3]$ that knows about the entire
moduli space of complex $G_{\C}$ flat connections (including abelian ones),
Seiberg-Witten invariants and Heegaard Floer homology.
Otherwise, there is no hope to reproduce Seiberg-Witten or Donaldson invariants \eqref{chiralcorr}
via gluing, if such information is already missing in $T[M_3]$.

\subsection{Orientation reversal}

The main theme of this article involves operations on algebras that mirror operations on 4-manifolds.
While we are mostly focused on operations of building 4-manifolds, {\it i.e.} various gluing constructions,
there are simple universal operations that do not involve cutting and gluing.
An example of such operation is the orientation reversal on the 4-manifold $M_4$:
\be
M_4 \quad \to \quad \bar M_4
\label{orientreverse}
\ee
It is natural to ask, then, whether there exists any relation between $\text{VOA} [\bar M_4]$ and $\text{VOA} [M_4]$.

The answer to this question is ``yes'' and turns out to be non-trivial already in the abelian case.
Indeed, even for $G=U(1)$ (and even more so for non-abelian $G$),
the left and right sectors of the 2d theory $T[M_4,G]$ are generically quite different,
as one can easily see from the values of $c_L (T[M_4,G])$ and $c_R (T[M_4,G])$.
The reason for this is that, even before dimensional reduction \eqref{CFTreduction},
the original theory in six dimensions is chiral.
And, the 2d theory $T[M_4]$ inherits this property, unless $M_4 = S^1 \times M_3$.

In particular, the number of right-moving and left-moving degrees of freedom in 2d theory $T[M_4]$
is determined by $b_2^+$ and $b_2^-$, respectively.
This is especially clear in the case of $G=U(1)$, where $b_2^+$ and $b_2^-$
tell us how many $\CN=(0,2)$ supermultiplets of different type compose the theory $T[M_4]$.
Because the orientation reversal \eqref{orientreverse} exchanges $b_2^+$ and $b_2^-$,
it also exchanges the two types of multiplets\footnote{Even though
for non-abelian $G$ the theory $T[M_4,G]$ is interacting,
it may still be helpful to think of it, roughly speaking, as consisting of three ``parts'' (for $b_1=0$):
two parts associated with $b_2^+$ and $b_2^-$, respectively, and another ``universal'' part
that is present for any $M_4$ and describes the center-of-mass degrees of freedom $\cong T[S^4]$.
While the latter remains invariant, the first two roughly exchange their roles under \eqref{orientreverse}.}
in the 2d $\CN=(0,2)$ theory $T[M_4,U(1)]$.

Now let us consider what happens at the level of $\text{VOA} [M_4]$.
Since $\text{VOA} [M_4]$ is the chiral left-moving algebra defined
by taking $\bar Q_+$ cohomology\footnote{where $\bar Q_+$ denotes
the supercharge of the $\CN=2$ supersymmetry in the rigth sector},
one might naively think that $\text{VOA} [M_4]$, in contrast to $T[M_4]$, is not sensitive to $b_2^+ (M_4)$ at all.
This, however, is not the case and again $G=U(1)$ serves as a good illustration.
In this case, the algebra $\text{VOA} [M_4]$ is basically a lattice algebra for
\be
L \; = \; H_2 (M_4,\Z)
\label{Llattice}
\ee
modulo an important detail related to the role of $H^{2,\pm} (M_4,\R)$.

First, for simplicity, let us assume that $M_4$ is negative definite and simply-connected,
and then apply \eqref{orientreverse}.
As discussed in the above, when $b_2^+ = 0$ and $G=U(1)$,
the theory $T[M_4,G]$ consists of $b_2^-$ compact left-moving bosons
and a universal center-of-mass multiplet $\Phi$ that will not play a role in the present discussion.
The OPE coefficients of the chiral left-moving bosons are determined by
the intersection form $Q: L \times L \to \Z$, which plays the role
of a ``generalized Cartan matrix'' for the lattice VOA $\CV_L$:
\be
\text{VOA} [M_4,U(1)] \; = \; \CV_L \otimes \text{VOA} [S^4]
\label{M4latticeVOA}
\ee
For example, when $M_4$ is defined by a plumbing graph, we have
\be
Q_{ij}=\left\{
\begin{array}{ll}
 1,& \text{if $i$ and $j$ are connected by an edge}, \\
 a_i, & \text{if } i=j, \\
0, & \text{otherwise}.
\end{array}
\right.
%\qquad i,j \in \text{Vertices of }\Gamma
\label{Qmatrix}
\ee

\begin{figure}[ht]
\centering
\includegraphics[scale=2.5]{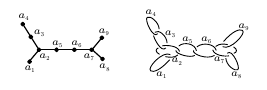}
\caption{Another example of a plumbing graph and the corresponding Kirby diagram.}
\label{fig:plumbing-example}
\end{figure}

If we apply \eqref{orientreverse}, we obtain a new 4-manifold $M_4$ with $b_2^- = 0$ and $b_2^+ > 0$.
The corresponding vertex algebra $\text{VOA}[M_4,U(1)]$ still has the form \eqref{M4latticeVOA},
except that now it is based on the lattice $(L,-Q)$ rather than $(L,Q)$.
This answer has two surprising features: one is the minus sign in $-Q$,
and the other is that $\text{VOA} [M_4]$ can ``see'' the self-dual part of $H^2 (M_4,\Z)$ at all.
After all, $H^{2,+} (M_4)$ governs the right-moving (supersymmetric) sector of the 2d $\CN=(0,2)$ theory,
not the left-moving (non-supersymmetric) sector.
These two features are, in fact, two sides of the same coin and have a common origin.

Once we apply \eqref{orientreverse}, we obtain a 2d theory with $b_2^+ = b_2$ compact right-moving bosons,
instead of $b_2^- = b_2$ compact left-moving bosons \cite{Dedushenko:2017tdw}.
Therefore, it appears that the lattice algebra $\CV_L$ is now part of the right-moving sector of $T[M_4]$.
While this is indeed correct, one should not forget that the right-moving $\CN=2$ supersymmetry pairs
all of these compact right-moving bosons $X_R$ with non-compact bosons $\sigma$ that have both left- and right-moving components.
In other words, the non-compact bosons $\sigma$ are {\it non-chiral} and, geometrically, they can be thought of
deformations of the coassociative submanifold \eqref{TM4def}, {\it i.e.} sections of $\Lambda^{2,+} (M_4)$.
As a result, BPS conformal fields $e^{i k_R \Phi} = e^{k_R (\sigma + i X_R) + \text{fermions}}$
contribute to the left-moving chiral $\text{VOA}[M_4]$ with an effective phase rotation by $i = e^{\frac{1}{2}\pi i}$
in the value of the right-moving momentum $k_R$ of the lattice VOA $\CV_L$.
This explains not only why $\text{VOA}[M_4]$ can ``see'' the right-moving sector associated with the self-dual
part of the lattice \eqref{Llattice}, but also why this contribution is accompanied by an extra minus sign
in the quadratic form $Q$.

In general, if $M_4$ is neither positive-definite nor negative-definite,
the BPS conformal fields that contribute to the left-moving chiral algebra $\text{VOA}[M_4, U(1)]$
have conformal weights
\be
h = \lambda_-^2 - \lambda_+^2 \,, \qquad \lambda:=(\lambda_-,\lambda_+) \; \in \; L^*
\label{confdimll}
\ee
where $L^*$ is the lattice dual to $L$,
and $\lambda_{\pm}$ denote projections of $\lambda$ to $H^{2,\pm} (M_4,\R)$
under a natural embedding $H^{2} (M_4,\Z) \subset H^{2} (M_4,\R)$.

\subsection{Chiral algebras and $\CN=(0,2)$ boundary conditions}

A plumbing graph, such as the one in Figure~\ref{fig:plumbing-example}, defines three manifolds at once:

$i)$ a 4-manifold $M_4$ with boundary,

$ii)$ its boundary $M_3 = \partial M_4$, and

$iii)$ a closed 4-manifold without boundary.

\noindent
The latter is (uniquely) constructed by attaching a 4-handle to $M_4$ which, in turn,
is constructed from the plumbing graph by attaching 2-handles to a single 0-handle according to the data of the plumbing graph.

Note, for $M_3$ and $M_4$ defined by the {\it same} plumbing graph,
the theories $T[M_4]$ and $T[M_3]$ are precisely in a relation illustrated in Figure~\ref{fig:M4d2d}.
Namely, 2d $\CN = (0,2)$ theory $T[M_4]$ defines a boundary condition for the 3d $\CN=2$ theory $T[M_3]$.
After the holomorphic twist \eqref{TM4def} we get, respectively,
a chiral algebra $\text{VOA} [M_4]$ and a modular tensor category $\text{MTC} [M_3]$,
both defined by the {\it same} plumbing graph and related as in \eqref{Krepring}.
This leads to a natural question, then:
If we are given a plumbing graph, like the one in Figure~\ref{fig:plumbing-example}, how do we construct a VOA and a MTC associated to it?

For the chiral algebra, we already saw the answer to this question in various examples
with $G=SU(2)$ and $G=U(1)$, see {\it e.g.} \eqref{A1toroidalY} and \eqref{M4latticeVOA}.
Now let us explain how to associate a modular tensor category to a general plumbing graph like
the one in Figure~\ref{fig:plumbing-example},
and then discuss its relation to the chiral algebra $\text{VOA} [M_4]$ defined by the same graph.

Since the general construction of Reshetikhin-Turaev uniquely defines a 3d TQFT from the data of
a modular tensor category, it is sometimes convenient to describe a MTC in terms of the corresponding 3d TQFT, from
which the modular tensor category can be extracted by asking what this 3d TQFT associates to a circle.
We will follow this strategy here, which will give us an easy way to access modular $S$ and $T$ matrices
of the MTC associated with a given (plumbing) graph.
The corresponding 3d TQFT can be constructed by assigning basic ingredients to edges and vertices
of the plumbing graph, according to the following rules:

\begin{itemize}

\item
{\bf Vertices:}
To a vertex with a framing coefficient $a_i \in \Z$ we assign a $U(1)$ Chern-Simons TQFT at level $a_i$:
\be
\overset{\displaystyle{a_i}}{\bullet}
\quad = \quad \frac{a_i}{4 \pi} \int A_i \wedge d A_i
\label{vertexrule}
\ee

\item
{\bf Edges:}
For each pair of vertices $(i,j)$ connected by an edge, we write a mixed Chern-Simons term:
\be
\frac{\phantom{xxxxx}}{\phantom{xxxxx}}
\quad = \quad \frac{1}{2 \pi} \int A_i \wedge d A_j
\label{edgerule}
\ee

\end{itemize}

\noindent
Putting all the contributions together, by summing over edges and vertices,
we obtain a 3d TQFT whose underlying MTC is the one we want, namely $\text{MTC} [M_3, U(1)]$.
For example, applying these rules to a graph in \eqref{MTCD4} we obtain a 3-fermion model, and so on.
The MTC associated to a graph according to these rules is invariant under operations
on graphs called 3d Kirby moves (see \cite{Gadde:2013sca} for a proof).

The full physical 3d $\CN=2$ theory $T[M_3]$ can be constructed by using a similar set of rules.
Namely, all we need to do is to supersymmetrize each of the ingredients \eqref{vertexrule} and \eqref{edgerule},
so that the resulting theory is a 3d $\CN=2$ supersymmetric quiver Chern-Simons theory
with level matrix \eqref{Qmatrix}, rather than the ordinary Chern-Simons TQFT.
It is amusing and instructive to verify that 3d $\CN=2$ theory constructed in this way
is also invariant under 3d Kirby moves.

As we explained earlier, in general, $T[M_4]$ associated with a plumbing graph
defines a 2d $\CN=(0,2)$ boundary condition in 3d $\CN=2$ theory $T[M_3]$ defined by the {\it same} graph.
In particular, it means that our 3d $\CN=2$ quiver Chern-Simons theories
come equipped with `canonical' BPS boundary conditions defined by the same plumbing graph.
Moreover, in the case of $G=U(1)$ and negative-definite matrix \eqref{Qmatrix},
we already know \eqref{M4latticeVOA} that the corresponding chiral algebra is
the lattice algebra $\CV_L$ based on the same quadratic form\footnote{Here and throughout the paper, we try to be careful distinguishing $H_2 (M_4,\Z)$, $H^2 (M_4, \Z)$, and other forms of cohomology. This is especially important when 4-manifolds in question have boundary, in particular in aspects related to cutting and gluing.
Note, the exact sequence for the pair $(M_4,M_3)$ reads in part
$$
\ldots \to H^1 (M_3,\Z) \to H^2 (M_4,M_3;\Z) \to H^2 (M_4,\Z) \to H^2 (M_3,\Z) \to H^3 (M_4,M_3;\Z) \to \ldots
$$
By Poincar\'e duality, $H^2 (M_4,M_3;\Z) = H_2 (M_4,\Z) = L$ and is the same as the second cohomology with compact support, $H^2_{\text{cpct}} (M_4,\Z)$. It is the lattice algebra $\CV_L$ for lattice $L$ (and not, say, $L^*$) that via Kaluza-Klein reduction \cite{Dedushenko:2017tdw} is identified with the $\text{VOA} [M_4]$ when $G=U(1)$, {\it cf.} \eqref{Llattice}--\eqref{M4latticeVOA} and section~\ref{sec:modules}. More precisely, $L$ is the free part $H_2 (M_4,\Z)$; the role played by the torsion part of $H^2 (M_4,\Z)$ and $H_2 (M_4,\Z)$ is discussed in \cite{Gadde:2013sca} and will not be directly relevant to us here. Therefore, for simplicity, we can tacitly assume that $\text{Tor} H^2 (M_4,\Z) = \text{Tor} H_2 (M_4,\Z) = 0$.} as the level matrix
in our 3d $\CN=2$ Chern-Simons theory $T[M_3]$.

An obvious choice of the `canonical' 2d $\CN=(0,2)$ boundary condition with these properties
%in the entire class of 3d $\CN=2$ theories defined by plumbing graphs
is the Dirichlet boundary condition for all 3d $\CN=2$ gauge multiplets.
As a check, one can consider the {\it half-index}
of the 2d-3d coupled system, first introduced in \cite{Gadde:2013wq},
{\it i.e.} $D^2 \times_q S^1$ partition function
of 3d $\CN=2$ theory with 2d $\CN=(0,2)$ boundary condition labeled by $\rho$ \cite{Gadde:2013sca}:
\be
Z_{T[M_3]} \left( D^2 \times_q S^1; \text{Dir} , \rho \right)
\; = \; \frac{\theta_{L+\rho} (q)}{\eta(q)^{b_2}}
\; \equiv \; \frac{1}{\eta(q)^{b_2}} \sum_{\lambda \in L + \rho} q^{\frac{1}{2} \lambda^T \cdot Q \cdot \lambda}
\label{ZMlattice}
\ee
Equivalently, this partition function can be described as the elliptic genus of the effective 3d $\CN=2$ theory $T[M_3]$ on $I \times T^2$ sandwiched by two $(0,2)$ boundary conditions, $\rho$ and Dir, at the two boundaries of the interval $I$.
In writing \eqref{ZMlattice}, we also used the identification between elements of the coset $L^*/L$ and characters of the lattice algebra $\CV_L$; it will be discussed further in section~\ref{sec:modules}, where notations such as ``$L+\rho$'' will be also explained.
According to \eqref{ZVWchar},
it should be equal to the Vafa-Witten partition function of $M_4$ with a boundary condition \eqref{rhorep} on $M_3 = \partial M_4$.
This is indeed the case. We will return to this point in \eqref{extendedchar} below.

This way of associating modular tensor categories to (plumbing) graphs can be generalized to higher-rank $G$.
While conceptually such constructions are very similar to the one presented here,
some of the details are quite subtle and will appear in our next paper.
For example, while a vertex of the graph also contributes to $T[M_3]$ a 3d gauge multiplet
with gauge group $G$ and $\CN=2$ Chern-Simons coupling at level $a_i$, {\it cf.} \eqref{vertexrule},
for general $G$ edges carry non-trivial matter charged under $G \times G$
and invariance of MTC under 3d Kirby moves becomes more interesting.

An important aspect of our discussion here was that a plumbing graph defines both 3d $\CN=2$ theory
and a 2d $\CN=(0,2)$ boundary condition in this theory.
Correspondingly, there are two types of dualities represented by operations on plumbing graphs:
those which don't change 3d $\CN=2$ theory with its MTC,
and those which keep invariant IR physics of 2d $\CN=(0,2)$ theory and its chiral algebra.
The former, called ``3d Kirby moves,'' are indeed the symmetries
of the above constructions and contain the latter, which will be our next topic.

\subsection{Simple Kirby moves}

Let us consider an example of gluing from \cite{Gadde:2013sca}
and see what happens at the level of chiral algebra $\text{VOA} [M_4]$.
In the notations of Figure~\ref{fig:M4d2d}, we take $M_4^+$ to be a linear plumbing with all Euler numbers $a_i = -2$,
{\it i.e.} an $A_{p-1}$ manifold \eqref{Apmanifold}:
\be
M_4^+ \quad = \quad A_{p-1}
\ee
The second part of our 4-manifold will be the total space of $\CO(-p)$ bundle over $\cp^1$:
\be
M_4^- \quad = \quad
\text{Tot}
\begin{bmatrix}
\CO (-p) \\
\downarrow \\
\cp^1
\end{bmatrix}
\quad = \quad
\overset{\displaystyle{-p}}{\bullet}
\ee
Both $M_4^+$ and $M_4^-$ have a Lens space boundary $L(p,1)$, in fact oriented just in the right way
so that $M_4^+$ and $M_4^-$ can be glued together along $M_3 = L(p,1)$ to form a closed 4-manifold without boundary.
Our first example of Kirby moves is the equivalence of $M_4 = M_4^+ \cup M_4^-$ and the connected sum of $p$ copies of $\bar \cp^2$,
\be
M_4 \quad = \quad
{\#}^p \, \left( \bar \cp^2 \right)
\quad = \quad
\underbrace{
\overset{\displaystyle{-1}}{\unknot}
\quad
\overset{\displaystyle{-1}}{\unknot}
\quad
\cdots
\quad
\overset{\displaystyle{-1}}{\unknot}
}_{p~\text{copies}}
\ee
which is a good illustration of basic handle slides~\cite{GompfS}.

Note, in the case $p=2$ all of the ingredients and the corresponding VOAs are especially simple.
Indeed, in this case we actually have $M_4^+ = M_4^- = A_1$, and the result of gluing is
a closed 4-manifold without boundary:
\be
M_4 \quad = \quad
A_1 \, \cup \, A_1
\quad \simeq \quad
{-1 \atop \unknot}
\phantom{mm}
{-1 \atop \unknot}
\ee
The corresponding vertex algebra clearly is
\be
\text{VOA} \big[ \; A_1 \, \cup \, A_1 \; \big]
~~=~~
\CU \otimes \CU \otimes \text{VOA} [S^4]
\ee
and the case of general $p$ is similar:
\be
\text{VOA} \Big[ \; \overset{\displaystyle{-p}}{\bullet} \, \cup \, A_{p-1} \; \Big]
~~=~~
\underbrace{
\CU \otimes \CU \otimes \cdots \otimes \CU}_{p~\text{copies}}
\otimes \text{VOA} [S^4]
\ee

Another simple example of a Kirby move is
\be
\overset{\displaystyle{-p}}{\bullet}
\frac{\phantom{xxxx}}{\phantom{xxxx}}
\overset{\displaystyle{-1}}{\bullet}
\quad \cong \quad
\overset{\displaystyle{-p+1}}{\bullet}
\qquad
\overset{\displaystyle{-1}}{\bullet}
\label{Kirbysimple}
\ee
which leads to the following equivalence of vertex algebras
\be
\text{VOA} \Big[ ~\overset{\displaystyle{-p}}{\bullet}
\frac{\phantom{xxxx}}{\phantom{xxxx}}
\overset{\displaystyle{-1}}{\bullet} ~ \Big]
\quad \cong \quad
\CU \otimes \text{VOA} \Big[ ~ \overset{\displaystyle{-p+1}}{\bullet}
%\qquad \overset{\displaystyle{-1}}{\bullet}
~ \Big]
\label{VOAsimple}
\ee
where both sides involve algebras from section~\ref{sec:extensions}.

\subsection{Triality}

A more non-trivial example of a Kirby move, which does not reduce to equivalences of quadratic forms,
is what in 4-manifold topology is known as a ``2-handle slide.''
Surprisingly, the corresponding equivalence of vertex algebras is not a duality, as one might naively expect,
but rather a triality symmetry \cite{Gadde:2013lxa}:
\be
\CT_{N_1, N_2, N_3} \; \cong \; \CT_{N_3, N_1, N_2} \; \cong \; \CT_{N_2, N_3, N_1}
\label{triality}
\ee
Here, $\CT_{N_1,N_2,N_3}$ is a vertex algebra that has three ``legs'' decorated, respectively,
by integers $N_1$, $N_2$, and $N_3$ that obey the ``triangle inequalities'', {\it i.e.} $N_2 + N_3 \ge N_1$, {\it etc.}
It can be defined as a chiral de Rham algebra of an (odd) vector bundler over
a Grassmannian.\footnote{Given a target manifold $X$,
possibly equipped with the odd bundle of left-moving fermions,
by de Rham chiral algebra we refer to the cohomology $H^* (X, \Omega_X^{\text{ch}})$
of the sheaf $\Omega_X^{\text{ch}}$. The latter is usually called the chiral de Rham complex of $X$.}

While Grassmannian varieties are ubiquitous in algebraic geometry and enjoy a well-known equivalence
\be
Gr(k,N) \; \cong \; Gr (N-k,N)
\label{Grbasic}
\ee
we are not aware of analogous trialities discussed in the algebraic geometry literature.
In fact, in order to upgrade the familiar duality \eqref{Grbasic} to a triality symmetry \eqref{triality}
it is crucial to make a few more key steps,
namely to equip the Grassmannian with an odd vector bundle and to pass to the chiral de Rham algebra.
Before we explain this, however, let us recall another basic fact, a description of the Grassmannian
$Gr (k,N) = \C^{kN} /\!\!/ U(k)$ as a K\"ahler quotient by $U(k)$. Then, from the viewpoint of such K\"ahler quotient,
the familiar equivalence \eqref{Grbasic} appears more non-trivial since it says
that $Gr (k,N)$ can be defined either by a system with $U(k)$ ``gauge group'' or, equivalently,
by a system with $U(N-k)$ ``gauge group.''

We can actually remove the quotes here by realizing such K\"ahler quotients in 2d quantum field theory,
where $U(k)$ and $U(N-k)$ indeed appear as gauge groups.
Moreover, we can try to make contact with VOAs (which describe chiral 2d conformal field theories)
by considering the chiral de Rham algebra of sigma-models with target spaces $Gr(k,N)$ and $Gr (N-k,N)$.
This almost works, and almost gives something interesting, but comes a little short:
mathematically, the resulting chiral de Rham algebra turns out to be quite trivial (does not lead to an interesting VOA)
and, physically, the reason is that 2d sigma-model with target space $Gr(k,N) \cong Gr (N-k,N)$ is a massive theory, not a CFT.

This last point has a simple fix, which takes us to a more interesting and non-trivial generalization of \eqref{Grbasic}
that {\it does} lead to equivalence of 2d CFTs, namely to the ``triality'' of the vertex operator algebras.
In order to describe it, let us introduce the universal (tautological) bundle $S$ associated
with the K\"ahler quotient $Gr (k,N) = \C^{kN} /\!\!/ U(k)$, and let $Q$ be the quotient bundle.
These bundles fit into the following short exact sequence \cite{Gadde:2014ppa,Sharpe:2015vza}:
\be
0 \; \longrightarrow \; S \; \longrightarrow \; \CO^N  \; \longrightarrow \; Q  \; \longrightarrow \; 0
\ee
Then, \eqref{Grbasic} becomes the following equivalence of the bundles
\be
S \; \to \; Gr (k,N) \qquad \cong \qquad
Q^* \; \to \; Gr (N-k, N)
\ee
and, conversely,
\be
Q \; \to \; Gr (k,N) \qquad \cong \qquad
S^* \; \to \; Gr (N-k, N)
\ee
Now we have enough equivalence relations to be assembled into a ``triality.''
In particular, if we define
\be
\CT_{N_1, N_2, N_3} \; := \; \text{de Rham chiral algebra}
\begin{pmatrix}
\Pi S^{\oplus N_3} \oplus \Pi Q^{\oplus N_2} \\
\downarrow \\
Gr \left( \frac{N_1 + N_2 - N_3}{2}, N_1 \right)
\end{pmatrix}
\label{TNNNGr}
\ee
where $\Pi S$ and $\Pi Q$ denote\footnote{In 2d $\CN=(0,2)$ superconformal theory,
these odd vector bundles are bundles of left-moving fermions.}
the shifted universal bundle and the shifted quotient bundle, respectively,
then $\CT_{N_1,N_2,N_3}$ enjoys \eqref{triality}.
Note, since the triple of integers $(N_1,N_2,N_3)$ obeys triangle inequalities,
the right-hand side of \eqref{TNNNGr} is well defined for any permutation of $(N_1,N_2,N_3)$.

Furthermore, the resulting vertex algebra $\CT_{N_1, N_2, N_3}$ contains three affine symmetries
$\hat{\frak{sl}(N_i)}_{n_i}$ at levels $n_i$:
\be
n_i \; := \; \frac{N_1 + N_2 + N_3}{2} - N_i
\qquad\qquad
(i=1,2,3)
\ee
and, in total, $\CT_{N_1, N_2, N_3}$ has the affine current algebra
\be
\bigotimes_{i=1}^3 \hat{\frak{sl}(N_i)}_{n_i} \otimes \hat{\frak{gl}(1)}_{\frac{N_i (N_1 + N_2 + N_3)}{2}}
\ee
with the Sugawara central charge
\be
c_L \; = \; \sum_{i=1}^3 \left( \frac{n_i (N_i^2 - 1)}{n_i + N_i} + 1\right)
\ee
Interpreting $\CT_{N_1, N_2, N_3}$ as a ``vertex'' with three legs, each carrying affine symmetry
$\hat{\frak{sl}(N_i)}_{n_i}$, $i=1,2,3$, one glue multiple copies of these basic VOAs
into larger quiver-like structures by applying level-rank duality to various legs.

%%%%%%%%%%%%%%%%%%%%%%%%%%%%%%%%%%%%%%%%%%%%%%%%%%%%%%%%%%%%%%%%%%%%%%%%%%%%%%%%%%%%

\section{Vertex algebras and trisections}
\label{sec:trisections}

In this paper we discuss operations on chiral algebras associated to various ways of cutting and gluing 4-manifolds.
These include traditional ways, such as handle decomposition and Kirby calculus, as well as novel techniques such as trisections,
that will be our subject in this section.

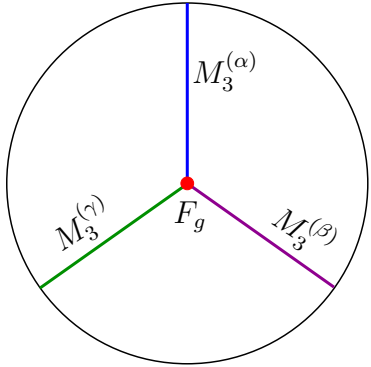
\begin{figure}[ht]
\centering
\bigskip
\bigskip
\bigskip
$\phantom{\int_{\int}}$\\
$\phantom{Z}$\\
\begin{picture}(100,100)
\put(-31,-10){\includegraphics[scale=0.3]{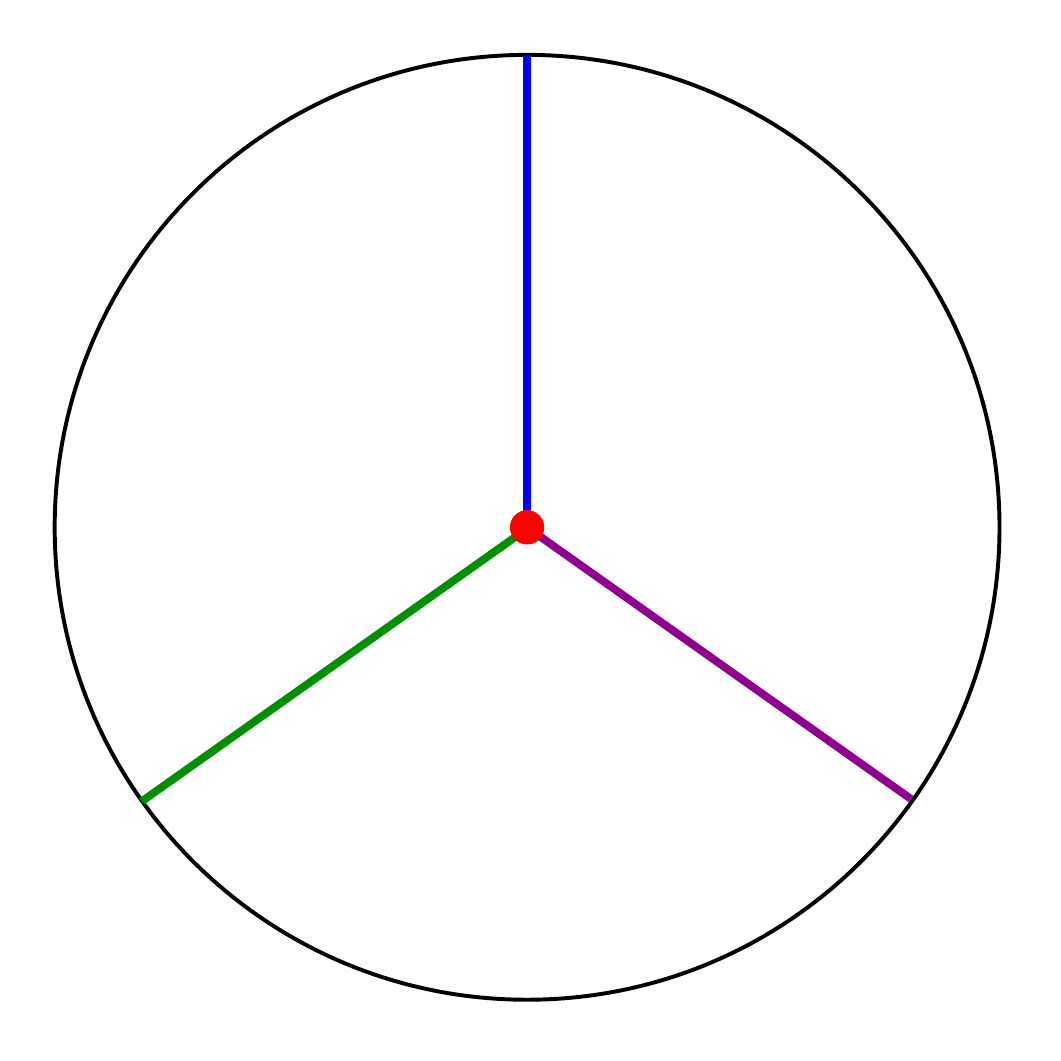}}
\put(40,52){{\Large $F_g$}}
\put(47,105){{\Large $M_3^{(\alpha)}$}}
\put(75,52){\rotatebox{-33}{{\Large $M_3^{(\beta)}$}}}
\put(-9,41){\rotatebox{33}{{\Large $M_3^{(\gamma)}$}}}
\end{picture}
\caption{A genus-$g$ trisection of a general smooth 4-manifold $M_4$ involves three 3d handlebodies
$M_3^{(i)} \cong \natural^g (S^1 \times B^2)$ and three 4d handlebodies $M_4^{(i)} \cong \natural^{k_i} (S^1 \times B^3)$.}
\label{fig:trisectMMM}
\end{figure}

Trisections of 4-manifolds, introduced by Gay and Kirby a couple of years ago \cite{MR3590351},
construct $M_4$ as a union of three four-dimensional handlebodies\footnote{More precisely, four-dimensional 1-handlebodies,
{\it i.e.} boundary connect sums $M_4^{(i)} \cong \natural^{k_i} (S^1 \times B^3)$.}
\be
M_4 \; = \; M_4^{(1)} \, \cup \, M_4^{(2)} \, \cup \, M_4^{(3)}
\label{MMM}
\ee
which pairwise meet along three-dimensional handlebodies $M_4^{(i)} \cap M_4^{(j)} \cong \natural^{g} (S^1 \times D^2)$ for $i \ne j$,
and such that the triple intersection
\be
F_g \; = \; M_4^{(1)} \, \cap \, M_4^{(2)} \, \cap \, M_4^{(3)}
\ee
is a closed orientable surface of genus $g$.
These pieces, as well as mutual relation between them, are illustrated in Figure~\ref{fig:trisectMMM}.
If a 4-manifold $M_4$ admits a $(g;k_1,k_2,k_3)$-trisection, then
\be
\chi (M_4) \; = \; 2 + g - k_1 - k_2 - k_3
\label{chigk}
\ee
Without loss of generality, one can focus on {\it balanced} trisections, {\it i.e.} trisections with $k_1 = k_2 = k_3$
(in which case we denote $k_i$ simply by $k$).
Then, $k \ge b_1 (M_4)$ and $g-k \ge b_2 (M_4)$ because $k$ determines
the number of 1-handles in a handle decomposition of $M_4$ and, similarly,
$g-k$ determines the number of 2-handles in the handle decomposition of $M_4$.
For example, these bounds imply that the minimal values of $g$ and $k$
are $(g,k) = (2 g_1 + 2, 2 g_1)$ for $M_4 = S^2 \times \Sigma_{g_1}$
and, similarly, $(g,k) = (4 g_1 g_2 + 2 g_1 + 2 g_2 + 2, 2 g_1 + 2 g_2)$
for $M_4 = \Sigma_{g_1} \times \Sigma_{g_2}$.

As usual, each genus-$g$ three-dimensional handlebody $M_4^{(i)} \cap M_4^{(j)}$, $i \ne j$,
can be conveniently described by a $g$-tuple $\{ \alpha_i \}_{1 \le i \le g}$ of closed curves on $F_g$.
In the present context, there are a total of three 3-dimensional handlebodies and,
correspondingly, three $g$-tuples of curves that traditionally are denoted by $\alpha$, $\beta$, and $\gamma$.
The data $(F_g;\alpha,\beta,\gamma)$, called the {\it trisection diagram} of $M_4$,
determines the homology and intersection form on a 4-manifold from the matrices
of intersection numbers $\langle [\alpha_i] , [\beta_j] \rangle_{F_g}$, {\it etc.}

\subsection{$\CN=(0,2)$ theories: 4d lift of 2d disk amplitudes}

\begin{figure}[ht]
\centering
\bigskip
\bigskip
\bigskip
$\phantom{\int_{\int}}$\\
$\phantom{Z}$\\
\begin{picture}(100,100)
\put(-31,-10){\includegraphics[scale=0.3]{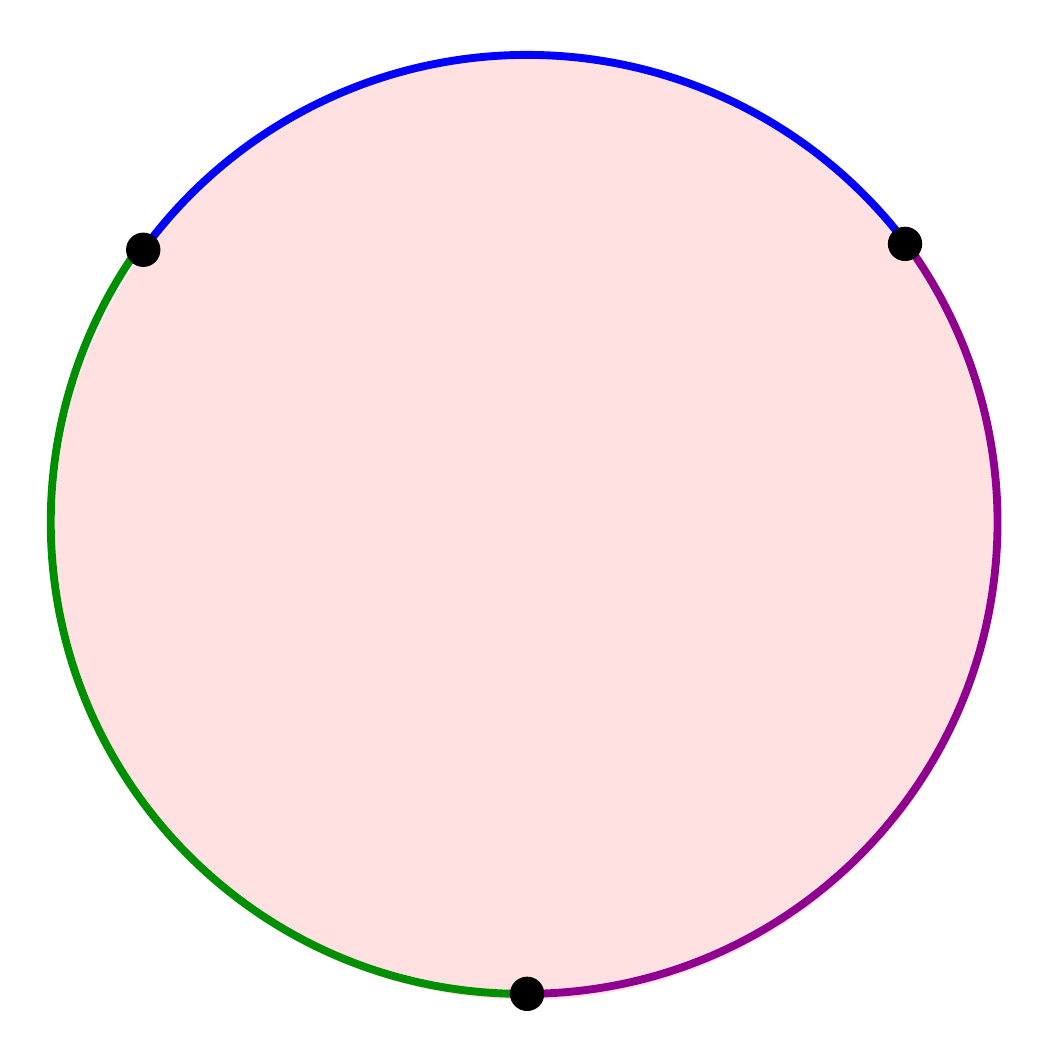}}
\put(37,140){{\huge $\CB_{\alpha}$}}
\put(113,40){{\huge $\CB_{\beta}$}}
\put(-40,40){{\huge $\CB_{\gamma}$}}
\put(60,47){{\Large $T (F_g)$}}
%\put(-23,20){\line(0,1){100}}
%\put(113,20){\line(0,1){100}}
%\put(-40,67){\line(1,0){180}}
%\put(-40,135){\line(1,0){180}}
%\put(0,-1){\line(1,0){100}}
\thicklines
\put(45,67){\circle*{5}}
%\put(45,67){\line(4,3){100}}
\multiput(45,67)(8,6){7}{\rotatebox{-52}{{\line(0,1){5}}}}
\multiput(40,67)(-8.1,6){7}{\rotatebox{52}{{\line(0,1){5}}}}
\put(8,100){\rotatebox{-37}{{\Large duality}}}
\put(3,83){\rotatebox{-37}{{\Large wall}}}
\multiput(45,-1)(0,10){7}{\line(0,1){5}}
\end{picture}
\caption{4d lift of 2d disk amplitudes from \cite{Gukov:2017zao}: namely, a
4d theory $T[F_g]$ on a disk with three
Heegaard boundary conditions $\CB_i$ and three 2d theories $V_i$ at the corners.
Dashed lines represent 3d duality walls which meet in the center of the disk.
The surface $\Sigma$ where $T[M_4]$ or $\text{VOA} [M_4]$ live
is orthogonal to the picture and not shown.}
\label{fig:diskBBB}
\end{figure}
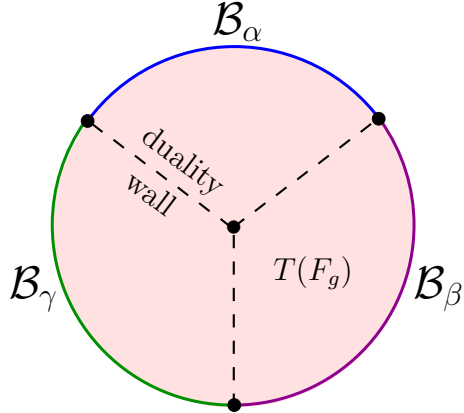

Each of the building blocks in a trisection of $M_4$
has a natural counterpart in the chiral algebra $\text{VOA} [M_4]$,
as summarized in Table~\ref{table:trisectVOA}.
Indeed, using the general rules from \cite{Gadde:2013sca} we see that
4d pieces $M_4^{(i)}$ correspond to vertex algebras $V^i$ that sit in corners
where two boundary conditions, $\CB_i$ and $\CB_{i+1}$, of class S theory $T[F_g]$ meet.
The $\text{VOA} [M_4]$ associated to a trisection is basically an extension of four VOAs at the corners of Figures \ref{fig:trisectMMM} and \ref{fig:diskBBB}.

Because $\CB_i$ and $\CB_{i+1}$ correspond to three-dimensional handlebodies
in the Heegaard splitting of a closed 3-manifold $\partial M_4^{(i)} \cong \#^k (S^1 \times S^2)$,
it is natural to call them {\it Heegaard boundary conditions}, {\it cf.} \cite{Gukov:2017zao}.
In fact, the setup in {\it loc. cit.} is related to our setup \eqref{TM4def}
here by exchanging the order of compactification on $\Sigma$ and $M_4$,
just as we did in our discussion above eq.\eqref{ZVWchar}.
Therefore, trisection of a non-Lagrangian 6d $(0,2)$ theory here should be viewed as a higher-dimensional
counterpart of a similar trisection of a non-Lagrangian 4d $\CN=2$ theory discussed in \cite{Gukov:2017zao},
class S theory $T[F_g]$ here should be viewed as a 4d analogue of the 2d theory $\CM (F_g)$,
and so on.

\begin{table}\begin{center}
\begin{tabular}{|c|c|}
\hline
\rule{0pt}{5mm}
Trisection of $M_4$ & Ingredient of $\text{VOA}[M_4]$ \\[3pt]
\hline
\hline
\rule{0pt}{5mm}
trisection surface $F_g$ & 4d $\CN=2$ theory $T[F_g]$ of class S \\[3pt]
\hline
\rule{0pt}{5mm}
3d handlebodies & 3d Heegaard boundary conditions \\[1pt]
$M_3^{(i)}$ & $\CB_i := \CB (M_3^{(i)})$ \\[3pt]
\hline
\rule{0pt}{5mm}
4d handlebodies & 2d $\CN=(0,2)$ interfaces $V_i$ \\[1pt]
$M_4^{(i)}$ & between $\CB_i$ and $\CB_{i+1}$ \\[3pt]
\hline
\end{tabular}\end{center}
\caption{The dictionary between geometry and algebra.}
\label{table:trisectVOA}
\end{table}

In order to construct the chiral algebra $\text{VOA} [M_4]$ from the building blocks
associated to a trisection of $M_4$, we need a good understanding of two key elements:
the Heegaard boundary conditions $\CB_i$ and the mapping class group $\text{MCG} (F_g)$
which acts by duality symmetries on 4d $\CN=2$ theory $T [F_g]$ of class S.
These two are not unrelated because all Heegaard boundary conditions
can be obtained from one basic Heegaard boundary condition $\CB_H$
via dualities
\be
\CB_{\varphi} \; = \; \varphi (\CB_H) \qquad , \qquad \varphi \in \text{MCG} (F_g)
\label{BviaBH}
\ee
In 4d theory $T[F_g]$, this operation can be understood as a process of colliding
a duality wall labeled by $\varphi$ with the basic boundary condition $\CB_H$.
And, this way of thinking about the Heegaard boundary conditions is especially convenient
for identifying the degrees of freedom on interfaces $V_{\varphi}$
(that were called $\CB_{\varphi}^{(2d)}$ in \cite{Gadde:2013wq}).
In other words, these are 2d degrees of freedom that one finds on a duality wall with boundary.

If we prefer to think in terms of elements of the mapping class (duality) group,
it is convenient to introduce $\varphi_{ij}$ labeled by pairs of indices $i \ne j$ that take values in a periodic set $\{ 1,2,3 \}$ (mod 3):
\be
\CB_i \; = \; \varphi_{ij} (\CB_j) \qquad , \qquad \varphi_{ij} \in \text{MCG} (F_g)
\ee
such that $\varphi_{ji} = \varphi_{ij}^{-1}$ and
\be
\varphi_{12} \circ
\varphi_{23} \circ
\varphi_{31} \; = \; \text{id}
\ee

Note, even in the basic abelian case --- that is, for $G=U(1)$ or $N=1$ in our notations ---
it is a fun and instructive exercise to derive the familiar 2d theories $T[M_4]$
and the corresponding VOAs by using this ``trisection approach.''
We leave it to an interested reader.

The next case of $N=2$ or $G=SU(2)$, which also happens to be our default choice for most examples
in this paper, is especially nice in the trisection approach since $T[F_g]$ admits
Lagrangian descriptions in different duality frames \cite{Gaiotto:2009we,Gaiotto:2009hg}.
This will be of great help in identifying the Heegaard boundary conditions $\CB_i$.

In the case of $N>2$ the resulting 4d $\CN=2$ theories are generically
non-Lagrangian.\footnote{One notable exception is $g=1$ which will be discussed separately below.}
While such theories are good candidates for producing new 4-manifold invariants,
unfortunately they are precisely the difficult ones from the viewpoint
of the trisection approach to $\text{VOA} [M_4]$
since duality walls and Heegaard boundary conditions are harder to describe in such theories.

There are several possible ways to tackle these challenges.
For example, one possibility --- already mentioned earlier --- is to exchange
the order of compactification on $\Sigma$ and $M_4$,
that is to start with the 4d $\CN=2$ theory $T[\Sigma]$
and then compactify it further on a 4-manifold $M_4$ defined by the trisection data.
The advantage of this approach is that the problem of defining Heegaard boundary conditions $\CB_i$
in a non-Lagrangian theory $T[F_g]$ is now reduced to a similar problem
in simpler 2d theory $\CM_{T[\Sigma]} (F_g)$.
The disadvantage, however, is that such method does not produce the entire $\text{VOA} [M_4]$,
but only computes its holomorphic partition function on $\Sigma$, as in \eqref{chiralcorr}.

Another way to say something concrete about the Heegaard boundary conditions
in a 4d non-Lagrangian theory $T[F_g]$ is to make use of the relation
\be
M_3 \; = \; M_3^{(i)} \cup_{F_g} M_3^{(j)} \; \cong \; \#^k (S^1 \times S^2)
\label{MijSS}
\ee
which must hold for all $i \ne j$.
According to the gluing rule analogous to that of Figure~\ref{fig:M4d2d},
this relation says that 3d $\CN=2$ theory $T[M_3]$ is equivalent (in the IR)
to a 4d theory $T[F_g]$ on an interval, with two Heegaard boundary conditions $\CB_i$ and $\CB_j$.
(If the starting point were in 4d rather than in 6d,
the analogous statement would be a version of the Atiyah-Floer conjectures
relating the Hilbert space of a 4d TQFT on $M_3$
to $\text{Hom} (\CB_{\alpha}, \CB_{\beta})$ in the A-model
of $\CM (F_g)$ with two branes $\CB_{\alpha}$ and $\CB_{\beta}$.)
Since the class of 3d $\CN=2$ theories $T[M_3]$ associated
by 3d-3d correspondence to 3-manifolds $M_3 = \#^k (S^1 \times S^2)$
is relatively well understood \cite{Gukov:2017kmk}, one might hope to use these relations
to constrain Heegaard boundary conditions $\CB_i$ in 4d theory $T[F_g]$.

\subsection{Simple examples: genus-1 trisections and corners}

When $g=1$, the corresponding 4d theory $T[F_g]$ is a maximally supersymmetric Yang-Mills theory with gauge group $G$.
It enjoys electric-magnetic duality symmetry $SL(2,\Z)$, which can be identified with the mapping class group of $F_g = T^2$.

Furthermore, for $g=1$ and arbitrary $G = U(N)$, the Heegaard boundary conditions
are easy to describe because, in this case, $M_3^{(i)} \cong S^1 \times D^2$.
Compactification of 6d $(0,2)$ theory on this 3-manifold with toral boundary
gives the basic Heegaard boundary condition in 4d $\CN=4$ super-Yang-Mills
that in \eqref{BviaBH} we denoted $\CB_H$.
Following the arguments of \cite{Nekrasov:2010ka}, it is easy to see that
$\CB_H$ corresponds to the principal Nahm pole boundary condition.
If we label it by $(1,0) \in H_1 (F_g)$, then S-duality \eqref{BviaBH}
maps it to another Heegaard boundary condition $\CB_{\alpha}$
labeled by $[\alpha] = (0,1) \in H_1 (F_g)$,
namely the Neumann boundary condition \cite{Gaiotto:2008ak}.

When a duality wall labeled by $\varphi \in \text{MCG} (F_g)$ runs to a boundary,
it changes the boundary condition\footnote{To be more precise,
the physical boundary condition stays the same, but its description changes,
in a way related to the original one by the duality $\varphi$.}
and, in general, leads to a 2d chiral algebra at the corner.
For $g=1$, simple examples of such 2d chiral algebras $V_{\varphi}$
were considered in \cite{Gadde:2013wq} (see also \cite{Gaiotto:2017euk}).
One crucial difference, however, is that in our present setup the superspace is
oriented a bit differently and, as a result,
any pair of Heegaard boundary conditions $\CB_i$ and $\CB_j$
preserves only 3d $\CN=2$ supersymmetry.
This is easy to see in the abelian case
and a detailed explanation for general $G$ can be found {\it e.g.} in \cite{Gukov:2017zao}.
For example, in our present setup, the boundary condition $\CB_H$ of Dirichlet type
freezes all fields on the boundary except for a single adjoint chiral multiplet
(that is usually called $\sigma$ in gauge theory approach to the geometric Langlands program, {\it cf.} \cite{Kapustin:2006pk}).

Unfortunately, the list of 4-manifolds that admit genus-1 trisections
is very short: $\cp^2$, $\bar \cp^2$, and $S^1 \times S^3$.
Since the first two already featured in our discussion, let us consider here
the remaining case of $M_4 = S^1 \times S^3$.
It has a very simple $(g,k)=(1,1)$ trisection diagram with $\alpha = \beta = \gamma$.
Therefore, it corresponds to 4d $\CN=4$ super-Yang-Mills on a disk, as in Figure~\ref{fig:diskBBB},
with no duality walls and the same boundary condition along the entire boundary of the disk,
$\CB_{\alpha} = \CB_{\beta} = \CB_{\gamma} = \CB_H$.
As we discussed a moment ago, this boundary condition supports a single adjoint chiral multiplet,
whose invariant polynomials generate the corresponding chiral algebra.
For $G=SU(2)$, there is only one invariant polynomial and, therefore, the chiral
algebra $\text{VOA} [M_4]$ consists of a single complex boson and its supersymmetric partner, a single complex fermion:
\be
\text{VOA} \big[ S^1 \times S^3 \big] \; = \;
\text{de Rham chiral algebra} \, \big( \Pi T^* \C \big)
\label{onechiral}
\ee
The corresponding (super)character in the RR sector has the form, {\it cf.} \cite{Gadde:2013wq}:
\be
%\chi_{\text{VOA}[S^1 \times S^3]} \; = \;
\chi (q,y) \; = \; \frac{\theta (y^{\frac{R}{2}-1};q)}{\theta (y^{\frac{R}{2}};q)}
\label{chiralindex}
\ee
where $R=4$, $y$ is the Jacobi variable, and
$\theta (x;q) = - i q^{\frac{1}{12}} x^{\frac{1}{2}} \prod_{n=0}^{\infty} (1 - x q^{n+1}) (1 - x^{-1} q^{n})$
is a ratio of the Jacobi theta-function $\theta_1 (x;q)$ and the Dedekind eta-function $\eta (q)$ that basically
equals the elliptic genus of a free $(0,2)$ Fermi multiplet.

%%%%%%%%%%%%%%%%%%%%%%%%%%%%%%%%%%%%%%%%%%%%%%%%%%%%%%%%%%%%%%%%%%%%%%%%%%%%%%%%%%%%

\section{Modules and knotted surfaces}
\label{sec:modules}

Our chiral algebras $\text{VOA} [M_4]$ have many interesting modules\footnote{Notice,
while rational VOAs may appear as intermediate building blocks in gluing constructions,
none of $\text{VOA} [M_4]$ is rational for a closed 4-manifold $M_4$.}
that correspond to supersymmetric (BPS) operators in 2d $\CN=(0,2)$ theory $T[M_4]$.

In particular, a large class of supersymmetric BPS operators in 2d comes from
supersymmetric BPS defects in 6d $(0,2)$ theory or, equivalently, from either M2 or M5-branes
in the full elevent-dimensional setup \eqref{TM4def}.
For such operators to be local ({\it i.e.} supported at points on $\Sigma$) and supersymmetric,
M2-branes (a.k.a. membranes) must be supported on associative submanifolds that meet $M_4$ along a surface $S$,
whereas M5-branes (a.k.a. fivebranes) must be supported on coassociative submanifolds in $\Lambda^{2,+} (M_4)$
and a fiber of $T^* \Sigma$.
The former are labeled by a vector in the weight lattice of $\frak{g} = \text{Lie} (G)$
and a choice of a 2-real-dimensional surface $S$, whereas the latter are labeled by
a nilpotent orbit of $\frak{g}$ or, equivalently, by a homomorphism $\frak{su}_2 \to \frak{g}$ (see {\it e.g.} \cite{Gukov:2006jk}).
Relegating the study of the latter to future work, here we focus on the former.

As a concrete example, let us consider the lattice VOA $\CV_L$ from \eqref{M4latticeVOA},
which is rational and packages an infinite set of Virasoro representations $V_{\lambda}$
%with characters $\chi (V_{\lambda}) = \frac{1}{\eta(q)^{b_2}} q^{\frac{1}{2} \lambda^T \cdot Q \cdot \lambda}$
into finitely many simple modules $V_{L + \rho}$ of the extended algebra labeled by $\rho \in L^* / L$,
where $L^*$ is the dual lattice \cite{DiFrancesco}.
The fusion product of these modules is simply
\be
V_{L+\rho} \times V_{L+\mu} \; = \; V_{L+\rho + \mu}
\ee
They have $q$-dimensions $\dim_q (V_{L+\rho}) = 1$ and characters
\be
\chi_{\rho} (q)
\; = \; \frac{\theta_{L+\rho} (q)}{\eta(q)^{b_2}}
\; \equiv \; \frac{1}{\eta(q)^{b_2}} \sum_{\lambda \in L + \rho} q^{\frac{1}{2} \lambda^T \cdot Q \cdot \lambda}
\label{extendedchar}
\ee
It is easy to verify that, when $M_4$ is a complex surface, this agrees with the counting
rank-1 sheaves\footnote{There is an important shift \cite{Gadde:2013sca} due to Freed-Witten anomaly that we suppress here.}
on $M_4$ or, equivalently, with the generating function of Euler characteristics
of Hilbert schemes of points on $M_4$,
\be
Z_{VW}^{U(1)} (M_4) \; = \; \sum_{m=0}^{\infty} \chi \left( \text{Hilb}^{m} (M_4) \right) \, q^m
\ee

This illustrates how winding-momentum conformal fields (or, states that they create from vacuum)
can be identified with M2-branes supported on embedded surfaces $S \subset M_4$.
In general, such surfaces can be rather complicated (knotted) but, at least for $G=U(1)$,
the weight $h (V_S)$ of the corresponding conformal field $V_S$
depends only on the homology class of $S$,
\be
\lambda \; = \; [S] \; \in \; L := H_2 (M_4,\Z)
\ee
In general, we expect the conformal weight $h (V_S)$ to be a quadratic function of $\lambda$, {\it cf.} \eqref{confdimll},
which on the one hand resembles the weight of a level-$N$ Kac-Moody representation, $h_{\lambda} = \frac{C_2 (\lambda)}{2 (N + h^{\vee})}$,
and on the other hand is related to the genus of the surface $S$, {\it cf.} \cite{Dedushenko:2017tdw}.

Both of these properties make $h (V_S)$ into a ``VOA counterpart'' of the so-called {\it genus function},
which to every class $\lambda \in L$ assigns a non-negative integer:
\be
G : \quad H_2 (M_4 , \Z) \; \to \; \Z_{\ge 0}
\ee
$$
\lambda \; \mapsto \; G(\lambda) = \text{min} \{ g(S) \; \vert \; S \subset M_4 \,, \; [S] = \lambda \}
$$
where $S$ ranges over closed, connected, oriented surfaces smoothly embedded in the 4-manifold $M_4$.
Note that $G(\lambda)$ is even, {\it i.e.} $G(- \lambda) = G(\lambda)$.
For example, for $M_4 = S^2 \times S^2$ we have (in the obvious basis):
\be
G : \quad (\lambda_1, \lambda_2) \; \mapsto \; (|\lambda_1|-1)(|\lambda_2|-1)
\ee
when $\lambda_1 \lambda_2 \ne 0$, and $G (\lambda) = 0$ otherwise \cite{Ruberman,Lawson}.
Similarly, for $M_4 = \cp^2 \# \bar{\cp}^2$,
\be
G: \quad \lambda_1 S_1 + \lambda_2 S_2 \; \mapsto \;
\begin{cases}
\frac{(|\lambda_1|-1)(|\lambda_1|-2)}{2} - \frac{|\lambda_2|(|\lambda_2|+1)}{2}, & \text{if } |\lambda_1| > |\lambda_2| \\
\frac{(|\lambda_2|-1)(|\lambda_2|-2)}{2} - \frac{|\lambda_1|(|\lambda_1|+1)}{2}, & \text{if } |\lambda_1| < |\lambda_2| \\
0, & \text{if } |\lambda_1| = |\lambda_2|
\end{cases}
\ee

\subsection{Wall crossing}

The above discussion also sheds light on a mysterious phenomenon of wall crossing.
Indeed, from the viewpoint of gauge theory on $M_4$, it is well known that for certain metrics on $M_4$
the space of solutions to partial differential equations like \eqref{HWpdes} or \eqref{NfSWeq} can jump,
and so the corresponding partition functions \eqref{ZVWchar} or \eqref{chiralcorr} as well as their
homological lifts {\it a la} \eqref{VOAmoduli} can all change.
This raises an obvious question: What is the 2d interpretation of such wall crossing phenomena in $T[M_4]$ or $\text{VOA} [M_4]$?

To answer this question, we first note that a prototypical case of wall crossing has to do with reducible solutions,
which sometimes can be negligible in partition functions, but leads to dominant contributions in homological invariants.
The simplest example of a reducible solution is an abelian flat connection on a $U(1)$ bundle $\CL \to M_4$ which satisfies $F^+ = 0$.
Clearly, for this to be the case, one needs to have $c_1 (\CL) \in H^2 (M_4,\Z) \cap H^{2,+} (M_4 , \R)$.
When the 4-manifold is negative-definite, this intersection contains the entire lattice \eqref{Llattice},
and so the $U(1)$ Vafa-Witten partition function \eqref{ZVWchar}
has the form of a character of extended algebra that we saw earlier in \eqref{extendedchar},
\be
\chi_{\rho} (q)
\; = \; Z_{VW} (M_4,\rho;q)
\; = \; \frac{\theta_{L+\rho} (q)}{\eta(q)^{\chi (M_4)}}
\ee
with the theta-function in the numerator.
In the opposite extreme case, when $M_4$ is positive-definite (or, more generally when $b_2^+ > 1$ and $M_4$ has generic metric),
the intersection in question contains only trivial homology class, and
\be
\chi_{\rho} (q)
\; = \; Z_{VW} (M_4,\rho;q)
\; = \; \frac{q^{\frac{1}{2} \langle \rho , \rho \rangle}}{\eta(q)^{\chi (M_4)}}
\ee
where we allowed for a possibility of a non-trivial boundary contribution $\text{CS} (\rho) = \frac{1}{2} \langle \rho , \rho \rangle$.
The key point is that the latter expression has no theta-function in the numerator.
{}From the gauge theory viewpoint, this is due to the fact that there are no reducible solutions.
And, from the viewpoint of chiral algebra, this is simply a character of $b_0 + b_2 + b_4$ free bosons
and $b_1 + b_3$ free fermions.

Based on this lesson, we propose the following interpretation to wall crossing phenomena:
\emph{in 2d conformal theory $T[M_4]$ they correspond to special values of the parameters such
that the chiral algebra $\text{VOA} [M_4]$ allows for further extensions.}

For example, in the case of a single compact boson on a circle of radius $R$
this can happen for rational values of $\frac{1}{2} R^2 \in \mathbb{Q}$.
This algebra is a (non-trivial) part of $\text{VOA} [M_4,U(1)]$ for
a 4-manifold $M_4 = S^2 \times S^2$, which has $b_2^+ = 1$ and indeed
exhibits wall crossing at these values of $\frac{1}{2} R^2 = \frac{\text{Vol} (S_1)}{\text{Vol} (S_2)}$.

\subsection{Bridge trisections}

It is a well known and widely used fact that 4d theories $T[F_g]$ of class S admit
supersymmetric (BPS) surface operators labeled by points of $F_g$ and weights of $\frak{g}$.\footnote{{\it e.g.} for $g=1$
it can be used to lift (categorify) the action of the affine Weyl group
on the cohomology (K-theory) of the moduli space of Higgs bundles with ramification
to the action of the affine braid group on the derived category of coherent sheaves on the same moduli space \cite{Gukov:2006jk}.}
In the context of AGT correspondence, such surface operators map to degenerate fields of $W$-algebras
and, in a similar setting of 3d-3d correspondence, they become line operators in $M_3$ as well as in 3d theory $T[M_3]$.
In particular, this correspondence applies to \eqref{MijSS} which describes 4d theory $T[F_g]$ on an interval,
with two Heegaard boundary conditions $\CB_i$ and $\CB_j$:
surface operators represented by points in $F_g$ extend to arcs inside each handlebody $M_3^{(i)}$.

These arcs in $M_3^{(i)}$ correspond to BPS line operators in 3d $\CN=2$ boundary theory $\CB_i$.
When we move from $\CB_j$ and $\CB_i$ (via $\varphi_{ij}$),
arcs become surfaces in $M_4$; they correspond to
line-changing operators at 2d $\CN=(0,2)$ interfaces $V_i$, {\it cf.} Table~\ref{table:trisectVOA}.
What we just described is the basic idea of bridge trisections \cite{MeierZupan},
translated into physics language of 2d-3d-4d coupled system illustrated in Figure~\ref{fig:diskBBB}.

Important to us here is that every knotted surface $S \subset M_4$ can be isotoped into a {\it bridge position}
with respect to the data of section~\ref{sec:trisections}, meaning that $S \cap M_3^{(i)}$
is a collection of trivial tangles and $S \cap M_4^{(i)}$ is a collection of unknotted disks.
Then,
\be
\chi (S) \; = \; - \frac{1}{2} \left| S \cap F_g \right| + \sum_{i=1}^3 \left| S \cap M_3^{(i)} \right|
\ee
Therefore, just like in \eqref{BviaBH} it was convenient to introduce
a basic Heegaard boundary condition $\CB_H$ associated to $\natural^{g} (S^1 \times D^2)$,
in our present discussion it is convenient to introduce a set of basic line operators in $\CB_H$ associated with trivial tangles
and, similarly, a collection of basic (``distinguished'') line-changing local operators at 2d $\CN=(0,2)$ interfaces.
In this setup, $\text{MCG} (F_g)$ is now enriched by the braiding operations.

To summarize, a knotted surface $S \subset M_4$ can be represented by a half-BPS surface operator
on (multiple cover of) a disk, same disk as shown in Figure~\ref{fig:diskBBB},
bounded by line operators on $\CB_i$.
For example, in the case of genus-1 trisections, these are half-BPS surface operators \cite{Gukov:2006jk}
that lead to modules of the vertex algebra $\text{VOA} [\cp^2]$ associated
with non-trivial knotted $S \cong \cp^1$ or a quadric curve.

%%%%%%%%%%%%%%%%%%%%%%%%%%%%%%%%%%%%%%%%%%%%%%%%%%%%%%%%%%%%%%%%%%%%%%%%%%%%%%%%%%%%

\section{Gluing via BRST reduction}

\subsection{Vertex algebras from $T[M_3]$}

We already used 3d-3d correspondence a number of times in our discussion, especially in questions
related to gluing 4-manifolds along 3-dimensional boundaries.
And, it can be of service to us once again, in computing $\text{VOA} [M_4]$ for
\be
M_4 \; = \; S^1 \times M_3
\label{M4M3}
\ee
where $M_3$ is an arbitrary 3-manifold.

While such non-simply-connected 4-manifolds can be subtle from other viewpoints,
they are actually easy to handle in the present approach since 2d theory $T[M_4]$
is essentially a dimensional reduction of the 3d theory $T[M_3]$.
Generally, the latter has $\CN=2$ supersymmetry, which upon dimensional reduction
gives rise to $\CN=(2,2)$ supersymmetry of 2d theory $T[M_4]$.
Furthermore, since we always define $T[M_4]$ as a conformal fixed point (in the deep IR),
the class of theories $T \big[ S^1 \times M_3 \big]$ comprises 2d $\CN=(2,2)$ super-conformal theories (SCFTs).
And, since $T[M_3]$ is known explicitly for large classes of 3-manifolds (Seifert manifolds, plumbings, {etc.}),
this provides a large supply of concrete examples where $\text{VOA} [M_4]$ can be written explicitly.

To summarize, for $M_4$ of the form \eqref{M4M3}, the full physical theory $T[M_4]$ has
not only the right-moving $\CN=2$ superconformal symmetry, but also left-moving $\CN=2$ superconformal symmetry.
This means that for any $M_4$ of the form \eqref{M4M3}, the corresponding $\text{VOA} [M_4]$
always has $\CN=2$ superconformal structure!

Prominent examples of such algebras include global cohomology of the chiral de Rham complex associated
to a Calabi-Yau manifold, see {\it e.g.} \eqref{onechiral} for one simple example.
In order to produce more examples, let us consider $M_3 = S^1 \times \Sigma_g$.
In this class of examples, the holonomy of the 4-manifold \eqref{M4M3} is reduced
even further and $\text{VOA} [M_4]$ has $\CN=4$ superconformal structure.
Specifically, it is a $\hat {\frak g}$-BRST reduction of $g+1$ symplectic bosons
and $g$ symplectic fermions, all valued in the adjoint representation of $G$:
\be
\text{VOA} \big[ T^2 \times \Sigma_g \big] \; = \;
%{\frak g}\text{-BRST reduction of } (\beta,\gamma)^{\otimes (g+1)} \otimes (b,c)^{\otimes (g+1)}
H_{BRST} \left( \big( {\frak g} \oplus {\frak g}^* \big)^{\otimes (g+1)}
\otimes \big( \Pi {\frak g} \oplus \Pi {\frak g}^* \big)^{\otimes g}
%\text{Bos} \big[ \text{Adj}^{\otimes (g+1)} \big]_{\mathbb{H}}
%\otimes \text{Ferm} \big[ \text{Adj}^{\otimes g} \big]_{\mathbb{H}}
\otimes (b,c)^{{\frak g}} \right)
\label{T2timesSigma}
\ee
Recall that, in general, the chiral algebra of a 2d vector multiplet is a $bc$ ghost system
valued in the adjoint representation of the gauge group $G$.
It has fermionic generators $b^A$ and $c_A$, $A = 1, \ldots, \dim G$,
\be
b^A (z) c_B (w) \; \sim \; \frac{\delta^A_B}{z-w}
\ee
and an affine $\hat{\frak{g}}$ current $J_{\text{gh}}^A = -i f^A_{BC} c^A b^B c^C$
at level $2 h^{\vee}$.
The chiral algebra of the gauged system, then, can be computed by the spectral sequence
applied to the $\bar Q_+$ cohomology, whose first page essentially implements
the Gauss law constraint:
\be
E_0 \; = \; \left( (\beta,\gamma) \otimes (b,c) \right)^G
\ee
The higher-order corrections to $\bar Q_+$ act on $E_0$ by the operator $Q_{BRST}$
constructed from the BRST current of the combined ghost-matter system,
\be
J_{BRST} \; = \; \sum_{A=1}^{\dim G} c^A \left( J_{\text{m}}^A + \frac{1}{2} J^A_{\text{gh}} \right)
\ee
It is easy to check that $Q_{BRST}$ is nilpotent, $Q_{BRST}^2=0$,
precisely when the levels of $\hat{\frak{g}}$ in the combined system add up to zero,
and so the perturbative chiral algebra of the system with gauged symmetry $G$
is given by the BRST reduction of the ungauged chiral algebra~\cite{Dedushenkonotes}:
\be
H_{\text{pert}} (\bar Q_+) \; \cong \; H_{BRST} (E_0)
\label{HBRST}
\ee
This construction appears in many examples of $\text{VOA} [M_4]$;
we already saw one class of examples in \eqref{T2timesSigma} and another one will be introduced shortly.

For $G=SU(2)$, the character of the algebra \eqref{T2timesSigma} is
\begin{multline}
\chi \; = \; \frac{1}{2} \frac{(q;q)_{\infty}^2}{\theta (y^{-1};q)} \oint \frac{dz}{2\pi i z}
\frac{\theta(z^{\pm 2}; q)}{\theta(z^{\pm 2} y^{-1}; q)}
\frac{\theta(z^{\pm 2} t; q)\theta(t; q)}{\theta(z^{\pm 2} ty; q)\theta(ty; q)} \times \\
\times \prod_{i=1}^g
\frac{\theta(u_i^{-1} z^{\pm 2} t^{-1} y^{-1}; q)\theta(u_i^{-1} t^{-1} y^{-1}; q)}{\theta(u_i^{-1} z^{\pm 2} t^{-1}; q)\theta(u_i^{-1} t^{-1}; q)}
\frac{\theta(u_i z^{\pm 2} y^{-1}; q)\theta(u_i y^{-1}; q)}{\theta(u_i z^{\pm 2}; q)\theta(u_i; q)}
%\frac{\theta (y^{\frac{R}{2}-1};q)}{\theta (y^{\frac{R}{2}};q)}
\end{multline}
where we use the shorthand notation $f (z^{\pm 2}) := f (z^2) f (z^{-2})$
and $\theta (x;q)$ is defined around \eqref{chiralindex}.
It would be interesting to compare the $u_i \to 1$ limit of this expression with
the equivariant elliptic genus of the moduli space of $SU(2)$ Higgs bundles on $\Sigma_g$.

\subsection{Product ruled surfaces}

Let $M_4 = \cp^1 \times \Sigma_{g,n}$, where $\Sigma_{g,n}$ is a two-real-dimensional surface of genus $g$ with $n$ boundary components.
As usual, $\Sigma_{g,n}$ can be constructed by gluing
$2g-2+n$ pairs-of-pants via $3g-3+n$ cylinders (or tubes):
\be
\# \Big( ~{\raisebox{-.6cm}{\includegraphics[width=1.6cm]{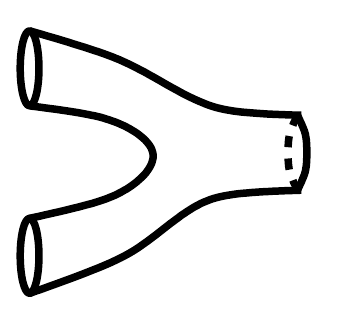}}} \Big) \; = \; 2g-2+n
\qquad , \qquad
\# \Big( ~{\raisebox{-.3cm}{\includegraphics[width=2.0cm]{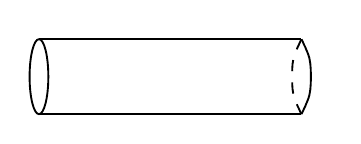}}} \Big) \; = \; 3g-3+n
\label{Sigmagntop}
\ee
and we are interested in the corresponding ``gluing'' at the level of $\text{VOA}[M_4]$.

Luckily, the full physical theory $T[M_4]$ for $M_4 = \cp^1 \times \Sigma_{g,n}$
has already been studied in \cite{Putrov:2015jpa}, from which one can deduce
the gluing rules for $\text{VOA}[M_4]$.
Note, since $M_4$ admits a metric of reduced holonomy, the supersymmetry of
the 2d theory $T[M_4]$ is enhanced from $\CN=(0,2)$ to $\CN=(0,4)$.
And, closely related to this is the fact --- already mentioned
in the footnote~\ref{foot:holonomy} --- that the central charge
in this case differs from the general formula \eqref{cLgeneric}.
In fact, the correct value of $c_L$ is quoted in Table~\ref{table:cLcR}.

The gluing rules are the following:
\be
\text{VOA} \Big[ \; {\raisebox{-.6cm}{\includegraphics[width=1.6cm]{pantsa}}} \Big] \; = \;
\begin{array}{l}
\beta \gamma \text{~system} \\
\text{associated to~} \C^2 \otimes \C^2 \otimes \C^2
\end{array}
\label{pantsbg}
\ee
\be
\text{VOA} \Big[ \; {\raisebox{-.3cm}{\includegraphics[width=2.0cm]{cylinder}}} \Big] \; = \; \text{BRST reduction with respect to}~ \hat{\frak{sl}(2)}
\label{tubegluing}
\ee
Specifically, to each copy of pairs-of-pants we associate 8 free complex bosons.
In the physical 2d theory $T[M_4]$ they come from 2d $\CN=(0,2)$ chiral multiplets
in the $({\bf 2}, {\bf 2}, {\bf 2})$ tri-fundamental representation
of $SU(2) \times SU(2) \times SU(2)$ flavor symmetry.
In particular, each copy contributes to the left and right central charges as follows:
\be
{\raisebox{-.6cm}{\includegraphics[width=1.6cm]{pantsa}}\,} : \qquad
c_L = 16
\qquad , \qquad
c_R = 24
\label{cLcRpants}
\ee
The corresponding chiral (left-moving) algebra involved in building $\text{VOA} [M_4]$
contains three copies of the affine
algebra $\hat{\frak{sl}(2)}$ at level $- \frac{1}{2} \times 4 = -2$.

Now, let us discuss gluing along the tubes \eqref{tubegluing}.
In the physical 2d theory $T[M_4]$ it corresponds to gauging $SU(2)$ flavor symmetry
associated with a given tube/cylinder.
In particular, it decreases the dimension of the target space
(when $T[M_4]$ can be thought of as a sigma-model) and lowers the values of central charges:
\be
\text{``gluing''} : \qquad
\Delta c_L = - 12
\qquad , \qquad
\Delta c_R = - 18
\label{cLcRtube}
\ee
Indeed, it was argued in \cite{Putrov:2015jpa} that the result of such
gluing operations applied to tri-fundamental $\CN=(0,2)$ chiral multiplets
is a sigma-model with target-space
$\mathbb{H}^{4(2g-2+n)} / \! / \! / \! / SU(2)^{3g-3+n}$
and a complex rank $2g$ bundle of left-moving fermions $E \to X$.
One important point in this construction, though, is that $SU(2)^{3g-3+n}$
does {\it not} act freely on $\mathbb{H}^{4(2g-2+n)}$ and, as a result,
the central charges quoted in Table~\ref{table:cLcR} are not simply
given by the sum of individual contributions \eqref{cLcRpants} and \eqref{cLcRtube},
counted with multiplicity \eqref{Sigmagntop}.

At the level of the vertex algebra that describes the left-moving chiral sector of $T[M_4]$,
the gauging of $SU(2)$ symmetries corresponds to the BRST reduction \eqref{HBRST},
where $G = SU(2)$ and $\hat{\frak{g}} = \hat{\frak{sl}(2)}$ or, rather, multiple copies of these.

In order for $\text{VOA} [M_4]$ to be a topological invariant of $M_4$,
the gluing rules in \eqref{pantsbg} and \eqref{tubegluing}
must satisfy relations that reflect different ways of constructing the same
$M_4 = \cp^1 \times \Sigma_{g,n}$.
As a good illustration of this,
suppose we glue two copies of the pairs-of-pants to produce $\Sigma_{0,4}$,
a genus-0 surface with four boundary components.
According to \eqref{tubegluing}, the corresponding vertex algebra $\text{VOA} [M_4]$
is a BRST reduction of two copies of the $\beta\gamma$ system \eqref{pantsbg},
and the 2d theory $T[M_4]$ is a $\CN=(0,4)$ super-QCD with $SU(2)$ gauge group and $N_f = 4$ flavors.
In particular, this result has manifest ``crossing symmetry,''
{\it i.e.} it is independent on how $\Sigma_{0,4}$ is glued out of
pairs-of-pants.

Note, that the six-dimensional setup relevant to our present discussion is $T^2 \times S^2 \times \Sigma_{g,n}$.
It is roughly a ``doubling'' of that in \cite{Beem:2013sza}, where chiral algebras are also associated to $\Sigma_{g,n}$ (and $\frak{g}$).
Correspondingly, the algebra $\text{VOA} \big[ S^2 \times \Sigma_{g,n} \big]$ is also roughly a doubling\footnote{We thank Pavel Putrov for useful discussions on this point.} of that in \cite{Beem:2013sza}.
Indeed, the basic building blocks \eqref{pantsbg} relevant to chiral algebras of \cite{Putrov:2015jpa} contain twice the number of generators compared to \cite{Beem:2013sza} and can be thought of as the ``complexification'' of the latter.
Similarly, if in the example of $\Sigma_{0,4}$ we replace $\CN=(0,4)$ super-QCD by a similar $\CN=(0,2)$ super-QCD with $SU(2)$ gauge group and $N_f = 4$ flavors, which has exactly half of the field content of the former, then the chiral algebra of the latter was shown \cite{Dedushenko:2017osi} to be $\hat{\frak{so}(8)}_{-2}$.
This relation between chiral of \cite{Putrov:2015jpa} and \cite{Beem:2013sza} is very intriguing and certainly deserves further study.

Moreover, we expect a similar relation between modular tensor categories, namely MTCs that the recent work \cite{Fredrickson:2017yka} associates to $\Sigma$ with a certain ramification data and $\text{MTC} [M_3]$ for $M_3 = S^1 \times \Sigma$, with the same $\Sigma$.
Indeed, the construction of MTC in \cite{Fredrickson:2017yka} is based on the geometry of the moduli space of Higgs bundles on $\Sigma$, which can be identified with the moduli space of complex flat connections,
$\CM_{\text{flat}} (G_{\C},\Sigma)$.
The geometry of the latter is precisely what enters the construction \cite{Gukov:2016gkn} of $\text{MTC} [M_3]$.

%%%%%%%%%%%%%%%%%%%%%%%%%%%%%%%%%%%%%%%%%%%%%%%%%%%%%%%%%%%%%%%%%%%%%%%%%%%%%%%%%%%%

\section{Associated variety of $\text{VOA}[M_4]$}

The notion of an {\it associated variety} of a highest weight module is a widely used and a very useful concept in representation theory; it offers a geometric look on representation theory, see {\it e.g.} \cite{Tanisaki,Vogan}.
Recently, a similar useful notion was introduced for vertex algebras \cite{Arakawa:2010},
and a natural question, then, is whether an associated variety of $\text{VOA} [M_4]$
has a nice physical interpretation in 2d theory $T[M_4]$.
(If so, it would provide yet another access to the structure of $\text{VOA} [M_4]$.)
Below we make a few general comments that we hope can be useful for addressing this question.

The first hint comes from the fact that
in many cases\footnote{The right technical term here is ``quasi-lisse,'' which means that
the associated variety of a strongly finitely generated VOA has finitely many symplectic leaves.}
the associated variety of a VOA carries a holomorphic symplectic structure.
While it plays a central role in matching the geometry of Higgs branches
in 4d $\CN=2$ theories \cite{Arakawa:2017aon}, this structure is way larger than what one should normally
expect in 2d $\CN=(0,2)$ theories, unless the supersymmetry is enhanced to $\CN=(0,4)$.
For theories $T[M_4]$ precisely this enhancement happens when a 4-manifold $M_4$ admits a K\"ahler metric,
and this therefore will be the context in which we expect to see a physical interpretation
of the associated variety of $\text{VOA} [M_4]$ in 2d theory $T[M_4]$.

Another useful hint comes from the fact that the associated variety is related to rationality of a VOA,
and often can be trivial (namely, a point) even if a VOA itself is interesting and non-trivial.
Indeed, a necessary condition for rationality of a VOA $\CV$ is that $\dim \CR_{\CV} < \infty$,
where $\CR_{\CV} = \CV / C_2 (\CV)$ is the so-called $C_2$-algebra of $\CV$.
And the same notion enters the definition of the associated variety:
\be
X_{\CV} \; := \; \text{Spec} (\CR_{\CV})_{\text{red}}
\ee
where $(\CR_{\CV})_{\text{red}}$ is the quotient of $\CR_{\CV}$ by its nilradical.
For example, $\CR_{\text{Vir}} = \C [x]$ for generic $c_L$
and $\CR_{\text{Vir}(p,q)} = \C [x] / (x^{\frac{1}{2}(p-1)(q-1)})$ for a $(p,q)$ minimal model,
so that $X_{\text{Vir}(p,q)} = \text{Spec} \C \cong \text{pt}$.
In a different example of $\CV = \hat{\frak{so}(8)}_{-2}$ mentioned earlier,
\be
X_{\CV} \; = \; \bar{\CO_{\text{min}} (D_4)}
\ee
is the closure of the minimal nilpotent orbit of $\frak{so}(8)$.
According to \cite{Putrov:2015jpa}, this is precisely the target space of the 2d theory $T[M_4]$ for $M_4 = \cp^1 \times \Sigma_{0,4}$ and $G=SU(2)$.
This suggests another way of expressing the relation between chiral algebras of \cite{Putrov:2015jpa} and \cite{Beem:2013sza} mentioned earlier; if we denote the latter by $\CV (\Sigma)$, then global sections of $\text{CDO} \big[ X_{\CV (\Sigma)} \big]$ provide a natural candidate for $\text{VOA} \big[ S^2 \times \Sigma_{g,n} \big]$, at least semi-classically.
We hope that future study of the associated variety of $\text{VOA}[M_4]$ can shed more light on this relation.

%%%%%%%%%%%%%%%%%%%%%%%%%%%%%%%%%%%%%%%%%%%%%%%%%%%%%%

\bigskip
{\it Acknowledgments~}
We would like to thank
A.~Braverman,
K.~Costello,
T.~Creutzig,
J.~Fuchs,
A.~Gadde,
J.~Meier,
H.~Nakajima,
N.~Paquette,
P.~Putrov,
M.~Rapcak,
C.~Schweigert,
J.~Teschner,
R.~Thomas,
and E.~Witten
for valuable discussions and inspiration,
F.~Ferrari and
S.~Lee
for comments on the manuscript.
We also wish to thank the anonymous referee for very useful feedback and many insightful comments.
The work of BF has been funded by the Russian Academic Excellence Project `5-100'.
Research of BF has also been supported by the Russian Science Foundation grant project 16-11-10316.
The work of SG is supported in part by
the U.S. Department of Energy, Office of Science, Office of High Energy Physics, under Award No. DE-SC0011632
and in part by the National Science Foundation under Grant No. NSF DMS 1664240.
This work is also supported in part by the ERC Starting Grant no. 335739 ``Quantum fields and knot homologies'' funded by the European Research Council under the European Union Seventh Framework Programme.

%%%%%%%%%%%%%%%%%%%%%%%%%%%%%%%%%%%%%%%%%%%%%%%%%%%%%%%%%%%%%%%%%%%%%%%%%%%

\bibliographystyle{amsalpha}

\end{document}